\documentclass[
 aip,
 amsmath,amssymb,
 reprint 
]{revtex4-1}

\usepackage{graphicx}
\usepackage{dcolumn}
\usepackage{bm}

\usepackage[utf8]{inputenc}
\usepackage[T1]{fontenc}
\usepackage{mathptmx}
\usepackage{etoolbox}

\usepackage{graphicx}
\usepackage{subcaption}

\usepackage{float}

\usepackage{comment}

\usepackage{color}
\usepackage{mathrsfs}

\makeatletter
\def\@email#1#2{%
 \endgroup
 \patchcmd{\titleblock@produce}
  {\frontmatter@RRAPformat}
  {\frontmatter@RRAPformat{\produce@RRAP{*#1\href{mailto:#2}{#2}}}\frontmatter@RRAPformat}
  {}{}
}%
\makeatother
\begin{document}

\preprint{AIP/123-QED}

\title{Autoregressive prediction of 2D MHD dynamics inferred from deep learning modeling}

\author{David Kivarkis}
  \email{david.kivarkis@etu.univ-amu.fr}
   \affiliation{Aix Marseille Université, CNRS, I2M, UMR 7373, 13331 Marseille, France}
   \affiliation{ECE (School of Engineering), Lyrids, Department of Applied Mathematics, 75015 Paris, France }

\author{Waleed Mouhali}
  \affiliation{ECE (School of Engineering), Lyrids, Department of Applied Mathematics, 75015 Paris, France } 
\author{Sadruddin Benkadda}

\affiliation{%
Aix Marseille Université, CNRS, PIIM, UMR 7345, 13397 Marseille, France }

\author{Kai Schneider}

\affiliation{Aix Marseille Université, CNRS, I2M, UMR 7373, 13331 Marseille, France}

\date{\today}

\begin{abstract}

We develop two deep learning surrogate autoregressive models for the prediction of the temporal evolution of two-dimensional ideal magnetohydrodynamic (MHD) Kelvin-Helmholtz instabilities across a range of magnetic field strengths. Using two neural network architectures, a Koopman-based Transformer model and a ConvLSTM-UNet, our approach enables simultaneous prediction of vorticity and current density directly from high-resolution simulations. The models are trained in an autoregressive manner and are able to reproduce key features of the multiscale dynamics over several instability growth and nonlinear saturation phases. Beyond accurate field reconstruction, the surrogates preserve essential physical structures of ideal MHD dynamics, including the conservation trends of global invariants and the propagation of Alfvénic fluctuations. Compared to direct numerical simulations, the proposed surrogates offer substantially reduced computational cost while maintaining good agreement with the reference dynamics. These results suggest that deep learning based surrogate models can provide a promising complementary tool for the efficient and physically consistent exploration of high-fidelity plasma and fluid simulations.\\

\end{abstract}

\maketitle

\section{\label{sec:Introduction}Introduction}

Understanding and forecasting the dynamics of magnetized flows remains a central challenge across plasma physics, astrophysics, and engineering applications such as fusion confinement and space propulsion~\cite{freidberg2014ideal}. The behavior of electrically conducting fluids under the influence of magnetic fields is governed by the equations of magnetohydrodynamics (MHD), which capture essential processes including instabilities, turbulence, and magnetic reconnection. These phenomena are not only fundamental to our theoretical understanding of plasmas but also critical for predictive modeling in practical systems. Traditional numerical solvers of the MHD equations yield high-fidelity results but incur significant computational cost, especially when high resolution or long forecasting horizons are required.
In response to these challenges, data-driven surrogate modeling and machine learning have emerged as promising tools for accelerating the prediction of nonlinear plasma and fluid dynamics~\cite{rosofsky2023magnetohydrodynamics, brunton2022data}. In magnetized plasmas, previous studies have shown that neural-network-based models can predict key turbulent transport quantities such as particle fluxes and Reynolds stresses with substantial computational speedups~\cite{varennes2025robust}, and have also been applied to edge-plasma tokamak modeling, efficiently predicting turbulent transport in fusion-relevant regimes~\cite{garrido2025ai}.

Beyond reduced transport models, deep learning has also been applied to forecast the spatio-temporal evolution of turbulent flow fields, predicting vorticity evolution in isotropic turbulence and reproducing essential turbulent structures~\cite{asaka2024machine}, as well as to Kelvin-Helmholtz instabilities in plasma systems using convolutional neural networks~\cite{hoshino2025estimation}.
More closely, in two-dimensional MHD flows, convolutional neural networks incorporating physical constraints have accurately predicted coupled velocity and magnetic fields for the canonical Orszag--Tang vortex problem~\cite{bormanis2024solving}.\\ 
New advances in deep learning have demonstrated the potential of autoregressive neural networks for forecasting high-dimensional turbulent flows directly from data. In particular, convolutional transformer architectures have been successfully applied to spatio-temporal prediction in two-dimensional hydrodynamic turbulence and wake flows \cite{patil2023autoregressive}. Recent work has also explored convolutional operator networks for spatio-temporal prediction in plasma turbulence, demonstrating both forward evolution and inverse parameter estimation for complex PDE systems such as the Hasegawa--Wakatani equations illustrated for low resolution simulation with considerable computational cost~\cite{chen2026convolution}. While this approach differs in physical system and goals, it exemplifies the growing application of deep learning surrogate models to multiscale physics problems.
These approaches leverage attention mechanisms to model nonlocal spatial interactions while performing multi-step autoregressive rollouts. However, their extension to magnetohydrodynamic (MHD) systems, especially in the context of plasma instabilities and coupled multi-field dynamics, remains largely unexplored, despite recent progress on Kelvin-Helmholtz instability modeling using physics-informed neural networks~\cite{wu2025physics}. In contrast to purely hydrodynamic settings, MHD flows exhibit additional nonlinear couplings and conservation constraints, posing significant challenges for data-driven forecasting models.\\
%\vspace{-0.1em}
In this work, we adress this gap by developping autoregressive deep learning architectures for forecasting the temporal evolution of vorticity and current density fields in two-dimensional incompressible ideal MHD, focusing on the Kelvin-Helmholtz instability as a representative nonlinear scenario. Specifically, we compare the Koopman Transformer~\cite{patil2023autoregressive} (KT-MHD) approach, which seeks to learn linearized dynamics in an embedded latent space, with a Convolutional Long Short Term Memory~\cite{desai2022next} (ConvLSTM) combined with U-Net~\cite{ronneberger2015u} model designed to capture spatio-temporal features directly. To the best of our knowledge, this study is among the first to benchmark these two architectures for MHD instability prediction and to analyze their predictive performance and generalization. Our results provide insights into the strengths and limitations of operator-inspired and recurrent convolutional approaches for data-driven MHD forecasting.\\
%\vspace{-0.1em}
This paper is organized as follows. Section \ref{sec:Ideal_Magnetohydrodynamics} introduces the physical context and governing equations. Section \ref{sec:Methodology} describes the numerical solver and dataset generation. Section \ref{sec:models} presents the deep learning architectures of two models, the Koopman Transformer (KT-MHD), and the ConvLSTM U-Net. Training procedures and evaluation metrics are detailed in Sections \ref{sec:Training} and \ref{sec:evaluation}. The analysis of the predictive performance for both models is discussed in Section \ref{sec:results}, followed by a summary and a discussion in Section \ref{sec:Conclusion}.

\section{Reduced Two-Dimensional MHD Model for the Kelvin-Helmholtz Instability}
\label{sec:Ideal_Magnetohydrodynamics}

Magnetohydrodynamics (MHD) describes the flow of electrically conducting fluids such as plasmas, liquid metals, and astrophysical flows by coupling the Navier–-Stokes equations with Maxwell's equations~\cite{freidberg2014ideal, biskamp2003magnetohydrodynamic}. 
In this framework, the flow velocity field $\mathbf{u}$ and magnetic field $\mathbf{B}$ are coupled: fluid motion stretches magnetic field lines, while magnetic tension and pressure influence the flow.
In the present work, we restrict our study to the ideal and incompressible MHD.

Under these assumptions, completed with suitable initial and boundary conditions, the governing equations are:
\begin{align}
\frac{\partial \mathbf{u}}{\partial t} + \mathbf{u} \cdot \nabla \mathbf{u} &= 
-\nabla p + (\nabla \times \mathbf{B}) \times \mathbf{B}, \label{eq:mom} \\
\frac{\partial \mathbf{B}}{\partial t} &= \nabla \times (\mathbf{u} \times \mathbf{B}), \label{eq:induction} \\
\nabla \cdot \mathbf{u} &= 0, \quad \nabla \cdot \mathbf{B} = 0. \label{eq:divs}
\end{align}
\noindent where $\nabla = (\partial_x, \partial_y)$ is the nabla operator and $p$ is the total pressure including magnetic contributions.

\subsection{\label{subsec:formulation_w_j}2D Formulation in terms of vorticity and current density}

In two-dimensions, the incompressible velocity and magnetic fields can be expressed by,

\begin{equation}
\mathbf{u} = \nabla^\perp \psi = (-\partial_y \psi, \partial_x \psi),
\qquad
\mathbf{B} = \nabla^\perp A = (-\partial_y A, \partial_x A),
\end{equation}
where $\psi$ is the stream function and $A$ the magnetic potential.

The vorticity and current density, which are both scalar-valued fields, are defined by:
\begin{equation}
\omega = (\nabla \times \mathbf{u})_z = \nabla^2 \psi, \qquad
j = (\nabla \times \mathbf{B})_z = \nabla^2 A.
\label{eq:poisson}
\end{equation}
where $\nabla^2$ denotes the Laplace operator.

The MHD system then reduces to two nonlinear transport equations:
\begin{align}
\frac{\partial \omega}{\partial t} + \mathbf{u} \cdot \nabla \omega &= 
\mathbf{B} \cdot \nabla j, \label{eq:vorticity}\\
\frac{\partial j}{\partial t} + \mathbf{u} \cdot \nabla j &= 
\mathbf{B} \cdot \nabla \omega. \label{eq:current}
\end{align}
Equations \eqref{eq:vorticity}--\eqref{eq:current} fully govern the system dynamics without explicitly evolving $\mathbf{u}$ and $\mathbf{B}$. 
The velocity and magnetic fields are reconstructed at each time step by solving Poisson equations for $\psi$ and $A$ (eqn.~\ref{eq:poisson}).\\

\subsection{\label{Kelvin-Helmholtz Instability}Kelvin-Helmholtz Instability}

We consider the MHD Kelvin-Helmholtz instability (KHI) which develops in shear flows and is strongly influenced by magnetic tension. It produces vortex roll-up, turbulent mixing, and strong coupling between hydrodynamic and magnetic structures~\cite{Miura1984KHI}.\\
This mechanism plays a fundamental role in mixing processes and turbulence generation in various astrophysical, geophysical~\cite{faganello2012magnetic}, and engineering contexts~\cite{Tirunagari2015KH}.
In plasma, this shear-driven instability is strongly modified by magnetic fields, making it a key benchmark to study transition to turbulence, mixing and vortex roll-up, and nonlinear coupling between hydrodynamic and magnetic structures.
Its rich spatial features and strong nonlinearities make KHI a highly relevant test case for data-driven surrogate models, especially when evaluating long-term prediction stability and physical consistency.

\section{Numerical Simulation of 2D Ideal Incompressible MHD}
\label{sec:Methodology}

\subsection{2D Incompressible Ideal MHD Solver}
\label{subsec:Solver}

The vorticity $\omega$ and current density $j$ are evolved in time and equations~\eqref{eq:vorticity}-\eqref{eq:current} are solved numerically using the Characteristic Mapping Method  \cite{yin2024characteristic} (CMM). 
This allows accurate tracking of nonlinear advection while preserving the divergence-free condition of both velocity and magnetic fields. 
The CMM was recently extended to handle non-linear advection problems with source terms, with a primary application to 2D ideal incompressible MHD. 
The solver employs a semi-Lagrangian formulation based on the evolution of the flow map (also called characteristic map) that is generated by the velocity field.
This approach enables thanks to compositional adaptivity high-resolution, non-diffusive simulations of ideal MHD turbulence.

A key feature of the method is its treatment of the Lorentz-force source term using Duhamel’s principles~\cite{courant2024methods}. The solution of the resulting inhomogeneous advection equation is expressed in terms of two quantities, the backward characteristic flow map, and the accumulated source term, defined as the Duhamel time integral of the Lorentz source.

Validation for the classical Orszag--Tang vortex benchmark~\cite{yin2024characteristic} confirms third-order accuracy in both space and time and demonstrates the method’s ability to capture thin current sheets and fine-scale features of ideal MHD turbulence. 
The resulting fields are therefore well-suited for machine learning applications, 
providing high-fidelity, low-diffusion datasets for data-driven modeling.
For more details on CMM, we refer to Yin et al.~\cite{yin2024characteristic}. 

From the characteristic mapping solver, the vorticity field $\omega$, encoding the rotational motion of the flow, and the current density $j$, characterizing the magnetic field gradients and current sheets, are computed and stored. The velocity field $\mathbf{u}$ is reconstructed from $\omega$ via the Biot-–Savart relation $\mathbf{u} = -\nabla \times \Delta^{-1}\omega$, and the magnetic field $\mathbf{B}$ is obtained from the current density $j$, correspondingly.

These scalar and vector fields form the basis of the machine learning dataset. In this study, the vorticity $\omega$ and current density $j$ are selected as the primary training and prediction variables, as they provide a compact yet dynamically rich representation of the coupled fluid–magnetic evolution.

\subsection{\label{subsec:Data Simulation}Generation of flow data}

The simulations are performed with CMM on a periodic two-dimensional domain 
${\cal D} = [0, L_x] \times [0, L_y]$ with $L_x = L_y = 2\pi$, 
%the Characteristic Mapping Method (CMM) 
to evolve the vorticity $\omega$ and current density $j$. 
The initial Kelvin-Helmholtz shear layer is prescribed through the vorticity profile
\begin{equation}
\omega(x,y,t=0)
= -\frac{\omega_0}{2a}
\left[ 1 - \tanh^2\!\left( \frac{\,y + \varepsilon \sin(x) - L_y/2\,}{a} \right) \right],
\label{eq:IC_vorticity}
\end{equation}
where  
$a = L_y/20$ defines the layer thickness, $\omega_0 = 1$, and $\varepsilon = 0.01$ controls 
the amplitude of the perturbation triggering the instability. 
This perturbed shear layer initiates the roll-up characteristic of the 
Kelvin-Helmholtz instability.\\
The magnetic field is initialized as a uniform background in the streamwise direction, i.e.,
$\mathbf{B}_0 = (B_{0x}, 0)$ 
so the initial current density is $j(x,y,0) = 0 $.

The simulation is sampled in time using a constant timestep $\Delta t_s$, defined as the temporal separation between two consecutive snapshots, in other terms, the snapshot time where the data are stored on the disk. 
In this work, we use $\Delta t_s = 0.5$ (non-dimensional time units), which ensures that the fast dynamics of the instability and nonlinear turbulence are properly captured.
This value refers to the output cadence only, not to the internal solver time step. 
The CMM integration itself uses an adaptive internal time step $\Delta t_{\mathrm{int}}$ (that varies from one time step to another), selected through a CFL (Courant-Friedrichs-Lewy) criterion~\cite{courant1928partiellen} and additional remapping imposed by accuracy constraints.
Because advection is treated along characteristic trajectories (a semi-Lagrangian formulation), 
the CMM is generally less restrictive than fully explicit time-stepping in Eulerian pseudo-spectral schemes, while still requiring practical step-size control to ensure accuracy. 
This separation between $\Delta t_{\mathrm{int}}$ and 
$\Delta t_s$ enables efficient data generation while preserving the fast Kelvin-Helmholtz growth and early nonlinear stage.

Simulations are performed for different initial uniform magnetic fields in streamwise direction:
\[
\mathbf{B}_0 = [0.05, 0], \; [0.08, 0], \; [0.1, 0], \; [0.12, 0].
\]

For the lower magnetic field case \(\mathbf{B}_0 = [0.05, 0]\), the simulation is run from \(t = 0\) to \(t_{end} = 50\), resulting in a total of 101 snapshots. For higher magnetic field strengths like \(\mathbf{B}_0 = [0.12, 0]\), the final simulation time is reduced (\(t_{end} = 30\)) to limit computational cost, as the simulations become more expensive for stronger fields. Increasing the magnetic field tends to stabilize the Kelvin-Helmholtz instability, slowing down the growth of vortices and turbulence~\cite{chandrasekhar2013hydrodynamic}.

The simulations are performed in the ideal MHD limit, the absence of explicit viscous and resistive dissipation (zero viscosity and zero resistivity) leads to the progressive formation of increasingly small spatial scales due to nonlinear vortex stretching and magnetic field line folding. As gradients intensify, thin current sheets and fine-scale structures develop, eventually approaching the grid resolution limit. This intrinsic scale contraction restricts the maximum reliable integration time of CMM since under-resolved gradients may degrade numerical accuracy and increase memory requirements.\\
For this reason, simulations are limited to approximately one characteristic vortex turnover time, ensuring that the dominant Kelvin-–Helmholtz roll-up and early nonlinear interactions are well resolved. Extending the simulations to significantly longer times would require higher resolution. The present study focuses on the ideal regime over a physically meaningful time horizon where the dynamics remain fully resolved.

\section{\label{sec:models}Deep Learning Architectures for MHD Forecasting}

This section presents two different approaches for MHD forecasting using deep learning. 
Subsection \ref{subsec:KT-MHD} %A
focuses on the Koopman transformer, discussing its self-attention mechanism, the Koopman operator, and the overall model architecture. 
Subsection~\ref{subsec:ConvLSTMUNet} %B 
introduces the ConvLSTM-UNet architecture, explaining the U-Net, the LSTM components, and their integration for spatio-temporal prediction.

\subsection{\label{subsec:KT-MHD}Koopman Transformer for MHD Spatio-Temporal Forecasting}

Recent advances in deep learning and surrogate modeling have enabled data-driven
forecasting of nonlinear dynamical systems at a fraction of the computational cost
of classical numerical solvers~\cite{karniadakis2021physics}.
Transformers have recently shown excellent performance for spatio-temporal forecasting
in fluid and plasma physics~\cite{patil2023autoregressive,long2025stft}. 

In this work, we introduce the Koopman Transformer for magnetohydrodynamics (KT-MHD), a hybrid architecture that combines convolutional encoders to extract spatial features from MHD fields, multi-head self-attention to model temporal dependencies, a recall Koopman operator theory to enforce a linear evolution in latent space, and a convolutional decoder to reconstruct the predicted physical fields from the latent space.

The goal of KT-MHD is to forecast the evolution of vorticity $\omega$ and current density $j$ from several previous simulation frames, using an autoregressive temporal rollout which will be explained in Section~\ref{sec:Training}.

\subsubsection{\label{subsec:attention}Self-Attention and Transformer Architecture}

Transformers, originally introduced by Vaswani et al.~\cite{Vaswani2017Attention},
were developed to model long-range dependencies in sequence processing tasks using self-attention mechanisms.
Unlike recurrent architectures (LSTM / ConvLSTM), the Transformer does not propagate information sequentially. Instead, it processes the full sequence in parallel and computes pairwise interactions between all temporal elements.

Given an input tensor $X \in \mathbb{R}^{T \times d}$, where $T$ corresponds to temporal frames and $d$ corresponds to the latent feature dimension used to represent each temporal frame,
self-attention computes three learned linear projections:
\[
Q = XZ_Q,\qquad K = XZ_K,\qquad V = XZ_V,
\]
where $Z_Q, Z_K \in \mathbb{R}^{d \times d_k}$ and $Z_V \in \mathbb{R}^{d \times d_v}$.
Thus, $Q, K \in \mathbb{R}^{T \times d_k}$ and $V \in \mathbb{R}^{T \times d_v}$.

Here, $d_k$ and $d_v$ denote the dimensionality of the query/key and value projections, respectively.
While $d_k$ controls the similarity computation between time steps, $d_v$ determines the amount of information carried by the attention mechanism.
In practice, $d_k$ and $d_v$ are typically chosen such that $d_k = d_v = d / n$, where $n$ denotes the number of attention heads in the Transformer block, ensuring that each attention head operates on a subspace of the latent representation.

The Attention operator is defined as:
\begin{equation}
\text{Attention}(Q, K, V)
= \text{softmax}\left( \frac{QK^\top}{\sqrt{d_k}} \right)V.
\label{eq:attention}
\end{equation}

The matrix product $QK^\top \in \mathbb{R}^{T \times T}$ encodes similarity scores between all pairs of time steps. 
More precisely, each element $(i,j)$ measures how strongly the $i$-th frame attends to the $j$-th frame.

The softmax function is applied row-wise, producing normalized attention weights that sum to one. 
This results in a data-driven weighting over all time steps, allowing each frame to aggregate information from the entire sequence.

Multi-head attention operates in parallel:
\begin{equation}
\text{MultiHead}(Q, K, V)
= \text{Concat}\big(\text{Attention}_1, \ldots, \text{Attention}_n \big)Z_O,
\end{equation}
where $n$ is the number of heads and $Z_O \in \mathbb{R}^{(n \cdot d_v) \times d}$ is a learnable projection that recombines the outputs of all attention heads and maps them back to the original feature dimension.

Each head learns distinct representations by projecting the input into different subspaces, enabling the model to capture complementary interactions at different spatial and temporal scales.

This property is particularly well suited to MHD turbulence, where coherent structures such as current sheets, vortices, and magnetic stretching interact across multiple scales in a highly nonlinear manner.

\begin{figure*}
\includegraphics[width=1\linewidth]{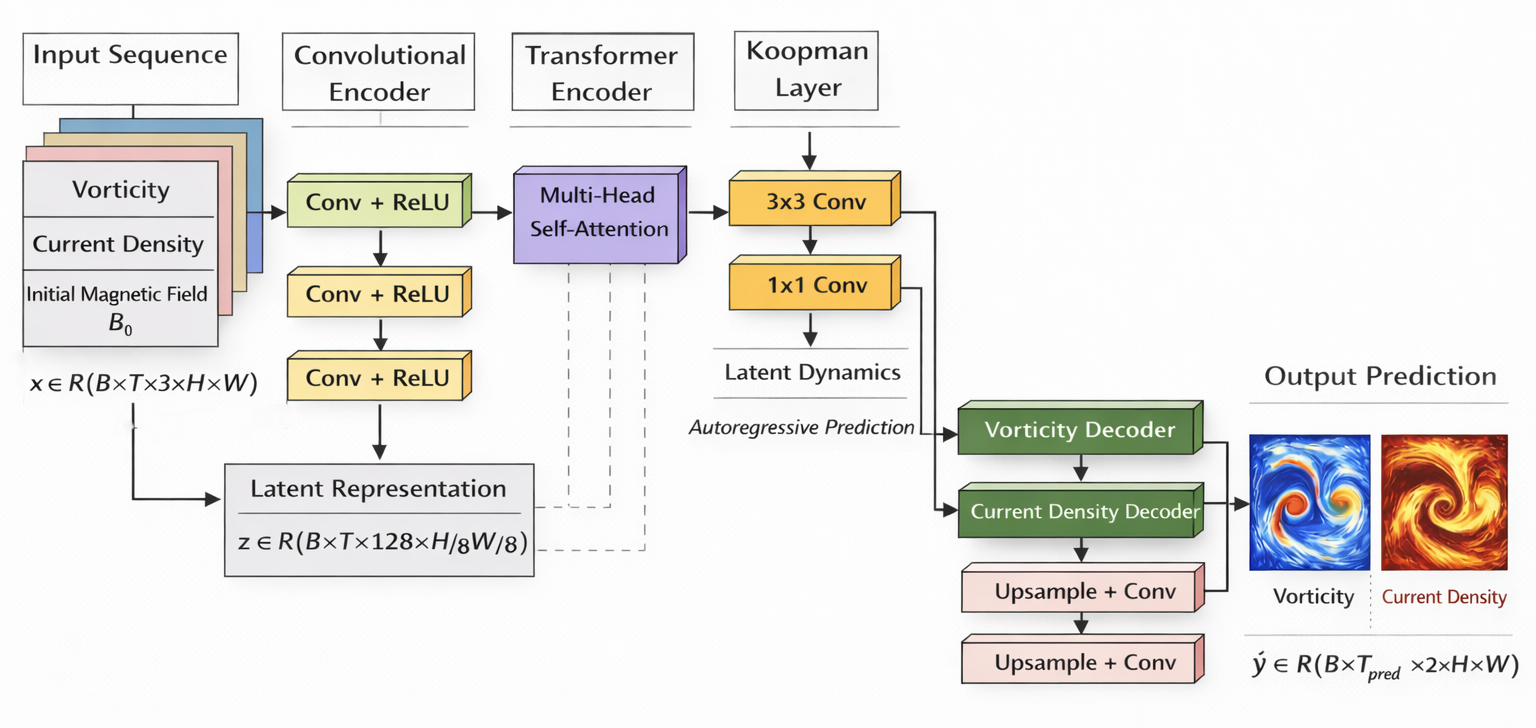}
\caption{\label{fig:KTMHD_architecture} Koopman Transformer architecture for spatio-temporal prediction of 2D incompressible MHD turbulence.}
\end{figure*}

\subsubsection{\label{subsec:koopman}Koopman Operator Theory and Latent Linear Dynamics}

We consider the discrete-time evolution of the incompressible 2D MHD system, where 
the physical state at time $t$ is represented by a field $x_t$. In our setting,
$x_t$ corresponds to the joint state $(\omega, j)$ where both vorticity and current density are predicted.

The nonlinear MHD dynamics can be written abstractly as
\begin{equation}
x_{t+1} = f(x_t),
\end{equation}
where $f$ denotes the nonlinear flow map of the ideal MHD equations. Building on Koopman operator theory~\cite{brunton2016koopman}, deep autoencoder architectures have been introduced
to learn nonlinear observables whose evolution is linear in a latent space, enabling
accurate and stable modeling of nonlinear dynamical systems~\cite{lusch2018deep, constante2024data}.

Koopman operator theory~\cite{brunton2019notes, brunton2022data} reformulates this nonlinear evolution by introducing
a nonlinear lifting map $g : \mathcal{X} \to \mathcal{Z}$, where  
$z_t = g(x_t)$ is an observable living in a (possibly infinite-dimensional) latent space.
In this lifted space, the evolution becomes linear under the Koopman operator $\mathcal{K}$:
\begin{equation}
\mathcal{K} g(x_t) = g(f(x_t)) = g(x_{t+1}).
\end{equation}

In practice, we seek for a finite-dimensional approximation of this operator.
Let $K \in \mathbb{R}^{m \times m}$ denote a learned linear map such that
\begin{equation}
z_{t+1} = K z_t,
\label{eq:koopman-finite}
\end{equation}
where $z_t = E(x_t)$ is obtained via a nonlinear encoder $E(\cdot)$.
A nonlinear decoder $D(\cdot)$ then maps the latent prediction back to physical space:
\begin{equation}
x_{t+1} = D(z_{t+1}) = D(K z_t).
\end{equation}

Thus the overall learned dynamical model takes the form
\[
x_t \xrightarrow{E} z_t \xrightarrow{K} z_{t+1} \xrightarrow{D} x_{t+1}.
\]

This decomposition allocates all nonlinearity to the encoder--decoder pair $(E,D)$,
while enforcing strictly linear temporal evolution in the latent space.
For MHD flows such as the Kelvin-Helmholtz instability, this structure provides enhanced long-term stability and significantly reduces error accumulation during autoregressive prediction.  
In particular, the latent linear operator $K$ acts as a finite-dimensional surrogate of the Koopman operator, enabling the model to capture coherent structures of the vorticity and current density fields through linear propagation in the lifted space.

\subsubsection{\label{KT_architecture}KT-MHD model Architecture}

The proposed model is a Koopman-based Transformer for coupled spatio-temporal forecasting of vorticity and current density in two-dimensional incompressible magnetohydrodynamics.

As shown in figure~\ref{fig:KTMHD_architecture}, the input consists of a temporal sequence
\begin{equation}
\mathbf{x} \in \mathbb{R}^{B \times T \times 3 \times H \times W},
\end{equation}
where $B$ is the batch size, $T$ is the temporal sequence length, the channel dimension $3$ corresponds to vorticity, the current density and the initial magnetic field, and $H$ and $W$ are the height and width of the spatial grid. The initial magnetic field magnitude $B_0$ is included as an additional channel to provide the model with information about the background magnetic field, which can influence the evolution of both vorticity and current density. Including $B_0$ allows the model to condition its predictions on different magnetic configurations without retraining separate networks. Each time slice is independently processed by a convolutional encoder composed of three convolutional layers. Each layer applies a $3\times3$ convolution with stride 2 followed by a ReLU activation (a simple thresholding nonlinearity defined as $f(x) = \max(0, x)$), resulting in progressive spatial downsampling and an increase in feature dimension up to a latent size of 128 channels. The encoder produces a latent representation
\begin{equation}
\mathbf{z} \in \mathbb{R}^{B \times T \times 128 \times H/8 \times W/8}.
\end{equation} 

Temporal correlations are modeled using a Transformer encoder operating exclusively along the time dimension. The Transformer consists of three stacked self-attention layers, each employing four attention heads. For each spatial location, the corresponding latent features across time are treated as a temporal sequence, allowing the model to capture long-range temporal dependencies while preserving spatial locality. Layer normalization is applied along the channel dimension of the latent tensor before feeding it into the Transformer. 

Time advancement is performed in latent space using a Koopman operator. The latent dynamics are governed by a single Koopman layer, implemented as a spatially local linear mapping composed of one $3\times3$ convolution followed by one $1\times1$ convolution for channel mixing. This operator is applied autoregressively to propagate the latent state over the prediction horizon. This design enforces a locally linear evolution in latent space while retaining spatial coupling, consistent with Koopman operator theory.

Reconstruction in physical space is carried out using two independent convolutional decoders, one for vorticity and one for current density. Each decoder consists of three convolutional layers, each followed by bilinear upsampling by a factor of two, and a final $3\times3$ convolution mapping the latent features to a single output channel. The outputs of both decoders are concatenated to form the final coupled prediction of vorticity and current density at full spatial resolution.
The final outputs are concatenated to produce a coupled prediction
\begin{equation}
\hat{\mathbf{y}} \in \mathbb{R}^{B \times T_{\text{pred}} \times 2 \times H \times W},
\end{equation}
where $B$ is the batch size, $T_{\text{pred}}$ is the prediction horizon, $2$ corresponds to the two physical variables (vorticity and current density), and $H$ and $W$ are the spatial height and width.

\subsection{\label{subsec:ConvLSTMUNet}ConvLSTM-UNet Architecture}

\begin{figure*}
\includegraphics[width=1\linewidth]{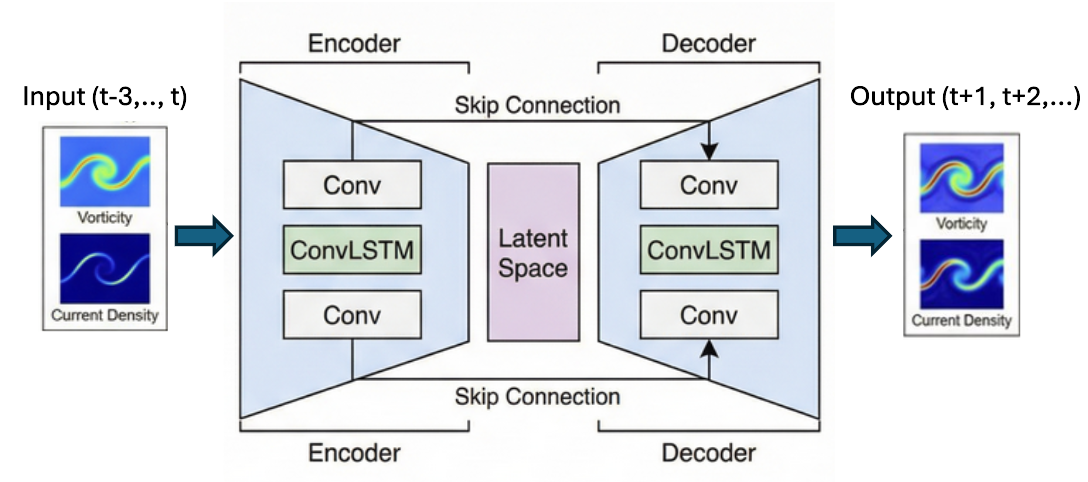}
\caption{\label{fig:lstm_unet_architecture} Diagram of the U-Net-ConvLSTM architecture.}
\end{figure*}

We introduce the ConvLSTM-UNet model shown in figure~\ref{fig:lstm_unet_architecture} , which is a combination of two architectures: U-Net and ConvLSTM. These components form the basis of the ConvLSTM-UNet, which will be detailed afterwards.

\subsubsection{\label{subsec:Unet}U-Net Encoder--Decoder Structure}

The U-Net architecture follows a symmetric encoder-decoder design that has proven highly effective for image-to-image regression and reconstruction tasks, making it well-suited for spatio-temporal fields such as those encountered in MHD simulations \cite{ronneberger2015u} .
The network is called a U-Net because its structure, when visualized, resembles the shape of the letter "U": the left side corresponds to the encoder path that progressively downscales the input, the right side corresponds to the decoder path that upsamples back to the original resolution, and the skip connections form horizontal links connecting the two sides, completing the "U" shape.

\textbf{Encoder:} The encoder consists of a series of convolutional blocks, each followed by a downsampling operation (typically max-pooling). At each stage, the network extracts increasingly abstract and high-level feature representations while progressively reducing the spatial resolution. This allows the model to capture large-scale structures and global context in the input data.

\textbf{Skip Connections:} To preserve fine-scale spatial information that may be lost during downsampling, U-Net employs skip connections that directly transfer feature maps from the encoder to the corresponding stage in the decoder. These connections help retain localized structures such as vortex filaments and current sheets, which are critical in MHD flows for accurately representing sharp gradients and small-scale interactions.

\textbf{Decoder:} The decoder mirrors the encoder in a symmetric fashion. It consists of upsampling operations (e.g., transposed convolutions) that progressively restore the spatial resolution of the feature maps. At each stage, the decoder concatenates the upsampled features with the corresponding encoder features through the skip connections, allowing the network to combine global contextual information with localized details. The final layer reconstructs the output field at the original input resolution, producing predictions that preserve both large-scale patterns and small-scale structures.

\subsubsection{\label{subsec:LSTM}ConvLSTM for spatio-temporal dynamics.}

Standard convolutional layers are well-suited for capturing spatial correlations in image-like data, but they do not account for temporal evolution. Conversely, standard LSTM networks are designed to model temporal dependencies but operate on vectorized inputs and therefore cannot exploit the spatial structure inherent in image or field data. 

ConvLSTM~\cite{desai2022next} layers combine the strengths of both approaches by replacing the fully-connected operations in LSTM cells with convolutions in both the hidden states and the gating mechanisms. This design enables the network to learn spatial correlations in $\mathbb{R}^{2}$ through convolutional filters, retain temporal memory across multiple prediction steps, and propagate motion and flow information through the hidden and cell states.\\

\subsubsection{\label{subsec:convLSTM_UNet}Final architecture : Combination of U-Net and ConvLSTM.}

The input tensor is defined as
\begin{equation}
\mathbf{x} \in \mathbb{R}^{B \times T \times 3 \times H \times W},
\end{equation}
where the three channels correspond to vorticity, current density, and the initial magnetic field $B_0$. Including $B_0$ as a conditioning channel allows the network to explicitly account for the influence of the mean magnetic field on MHD dynamics.\\
The encoder consists of three sequential encoder blocks, each comprising a 2D convolution followed by ReLU activation for spatial feature extraction, a max pooling for progressive downsampling and a ConvLSTM layer applied along the temporal dimension to capture short-term temporal dependencies.\\
The hidden dimensions of the encoder layers are $(32,64,128)$, providing a hierarchical representation from fine to coarse scales. Using a shared encoder for all input channels allows the network to learn cross-variable correlations in a fused latent representation.\\
At the bottom of the U-Net, a ConvLSTM layer with 128 channels operates on the coarsest spatial resolution, integrating temporal information across the full sequence. This bottleneck enables the network to propagate long-range temporal correlations, complementing the multi-scale spatial features extracted by the encoder.\\
The decoder mirrors the encoder and consists of three decoder blocks. Each block performs a transposed convolution (upsampling) to restore spatial resolution, a concatenation with the corresponding encoder skip connection, preserving high-resolution spatial information and a ConvLSTM applied on the upsampled feature sequence to refine temporal dynamics.\\
Skip connections are critical to maintain fine-scale spatial structures such as vortex cores and current sheets.
A final $3 \times 3$ convolution maps the last decoder features to two output channels corresponding to $\omega$ and $j$ fields. For multi-step prediction over a horizon h, the output $\hat{\mathbf{y}}$ is autoregressively rolled out using predicted sequences:
\begin{equation}
\hat{\mathbf{y}} \in \mathbb{R}^{B \times h \times 2 \times H \times W}.
\end{equation}

\vspace{-0.5em}

This architecture combines spatial feature extraction with temporal memory, making it well suited for nonlinear dynamical systems~\cite{li2024deep}. This capability is particularly relevant for forecasting the evolution of fluid and magnetic structures in 2D ideal MHD, where the dynamics are governed by coherent advection, vortex interactions, and strong nonlinear couplings. By jointly preserving spatial structure and temporal dependencies, ConvLSTM-based models provide a natural framework for representing complex spatio-temporal fields and chaotic dynamics~\cite{vlachas2018data}, enabling accurate autoregressive predictions in plasma simulations.

\begin{figure*}
\includegraphics[width=1\linewidth]{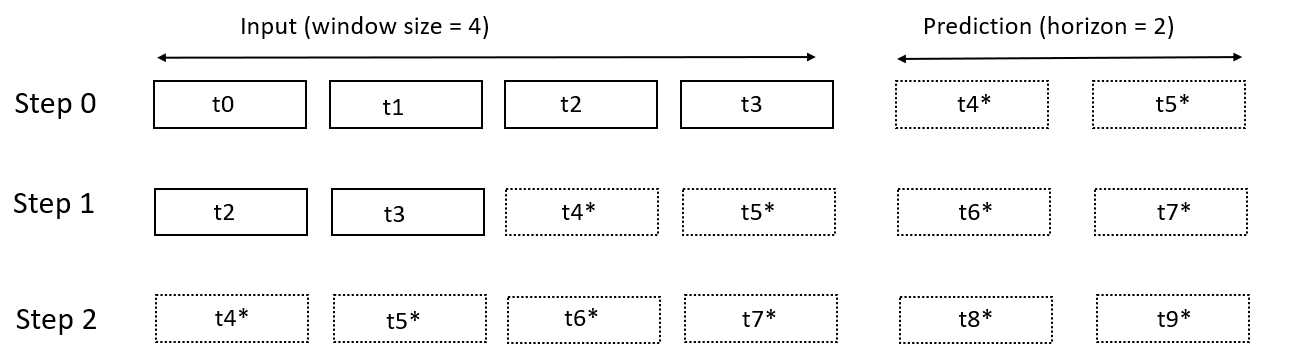}
\caption{\label{fig:Sequence_evolution} Illustration of the autoregressive prediction process. The stars ($*$) indicates predicted frames that are injected back into the sequence to generate further predictions.}
\end{figure*}

\section{\label{sec:Training}Learning Framework for Spatio-Temporal Prediction}

Both KT-MHD and ConvLSTM-UNet architectures are trained in an autoregressive manner on the same MHD dataset, where each forecasted frame is recycled as input for subsequent predictions. This strategy allows the model to learn long-term consistency and stability in MHD dynamics forecasting. To ensure fair comparison, both models are evaluated under identical forecasting strategies, described next.

\subsection{\label{subsec:prediction_strategies}Autoregressive Strategy}

As shown in figure \ref{fig:Sequence_evolution}, we take an initial sequence of four frames and predict the next two frames. The predicted frames (denoted with a star, e.g., t4*) are  injected back into the input sequence to form the next input, and this process is repeated several times.\\
Rather than fixing the number of input frames, predicted frames, and autoregressive iterations manually, we let the model discover the best configuration. These quantities are optimized via Bayesian optimization with Optuna, described in the following section.

\subsection{\label{sec:Optuna}Hyperparameter Optimization with Optuna}

To optimize the performance of the two architectures considered in this work, we employ Optuna~\cite{akiba2019optuna}, an open-source hyperparameter optimization framework. Optuna utilizes a flexible and efficient search algorithm based on Bayesian optimization with pruning strategies to identify optimal hyperparameter configurations. An objective function was defined to evaluate the model's validation loss for a given set of hyperparameters. The search space is explored using the Tree-structured Parzen Estimator (TPE), and unpromising trials are dynamically pruned to save computational resources.

The hyperparameters tuned with Optuna, along with their search ranges, are summarized in Table~\ref{tab:tuned_params}.

\begin{table}[h]
\centering
\caption{Hyperparameters search range.}
\label{tab:tuned_params}
\begin{tabular}{ll}
\hline
\textbf{Parameter} & \textbf{Range} \\
\hline
Window size & $[4,10]$ \\
Horizon & $[2,5]$ \\
Rollout & $[3,6]$ \\
Learning rate & $\{10^{-5}, 3\cdot10^{-5}, 10^{-4}, 3\cdot10^{-4}, 10^{-3}\}$ \\
Batch size & 2 \\
Latent dimension & 128 \\
\hline
\end{tabular}
\end{table}

The window size, which represents the number of consecutive input frames used for prediction, controls the temporal context provided to the model, with larger windows capturing more dynamics at the cost of increased memory usage. The horizon determines how many future frames are generated in a single forward pass, while the rollout defines the number of autoregressive steps, thereby controlling the total forecasting length. Finally, the learning rate governs the optimization process, balancing convergence speed and stability. Certain parameters such as the batch size and the latent dimension were held fixed to ensure compatibility with GPU memory constraints and training stability. These choices reflect a trade-off between model complexity and hardware limitations, ensuring efficient training without compromising predictive performance.

These parameters were explored across 50 trials, each training the model for 5 epochs in order to balance computational cost and model performance. The best combination was selected based on the lowest validation Mean Squared Error (MSE).\\
Based on the outcomes of the 50 trials, the configuration yielding the lowest validation MSE was selected for training. This optimal setup : window size $=4$, horizon $=2$, rollout $=3$, and learning rate $=10^{-4}$ offered the best trade-off between predictive accuracy and computational cost.\\
The Optuna optimization consistently identified a rollout value of $R=3$ as the optimal choice, yielding excellent predictive accuracy and stable autoregressive behavior. However, while this configuration provides highly accurate short-term forecasts, it limits the temporal extent of the predictions.\\
To assess the robustness of the model and its ability to extrapolate further in time, we additionally evaluated a larger rollout value of $R=6$, which enables longer-term forecasting. Despite the increased difficulty of long-term forecasting, this extended rollout configuration also produced stable and accurate predictions, and its qualitative and quantitative performance is discussed in the following sections.\\
We further investigated larger rollout values up to $R=8$. For rollouts beyond $R=6$, the model exhibits a gradual degradation in predictive accuracy together with increasing instability in the autoregressive predictions. Although the model still reproduces the overall dynamics, these results highlight the inherent limitations of the current architecture for extended long-term forecasting.

\subsection{Data Processing}
\label{subsec:data}

To assess the ability of the models to generalize across different magnetic field strengths, the training and testing datasets are constructed using distinct values of the imposed uniform magnetic field $B_0$.

Specifically, both the KT-MHD Transformer and the ConvLSTM-UNet models are trained using simulations corresponding to
\[
B_0 = [0.05, 0] \quad \text{and} \quad B_0 = [0.10, 0].
\]
These cases are used to learn the coupled spatio-temporal dynamics of vorticity and current density under different levels of magnetic influence.

Model evaluation is then performed on simulations with two unseen magnetic field strengths,
\[
B_0 = [0.08, 0] \quad \text{and} \quad B_0 = [0.12, 0],
\]
which are not included during training. The case $B_0 = [0.08, 0]$ corresponds to an interpolation between the trained regime, while $B_0 = [0.12, 0]$ represents an extrapolation to stronger magnetic fields.

This setup enables a systematic evaluation of the models' generalization capabilities with respect to changes in the background magnetic field, a critical requirement for data-driven forecasting of MHD turbulence.

To prepare the training data, overlapping input–output pairs are extracted using a sliding window approach. Each sample consists of a short history of four consecutive frames $(t-3, t-2, t-1, t)$, corresponding to an input window size of 4, which is used to predict the two future states $(t+1, t+2)$, defining thus a prediction horizon of 2. The total number of available frames is denoted by $N$, corresponding to all snapshots aggregated from the datasets with initial magnetic fields $B_0 = [0.05, 0]$ and $B_0 = [0.10, 0]$.\\

The number of available training sequences extracted from a dataset is given by :

\begin{equation}
N_{\text{seq}} = N - W - (R \cdot H) + 1,
\end{equation}
where $W$ denotes the input window length, $H$ the prediction horizon, and $R$ the number of autoregressive rollouts.\\

In our configuration ($W=4$, $H=2$, $R=6$), this yields for the dataset with $\mathbf{B}_0=[0.05,0]$
\begin{equation}
N_{\text{seq}}^{(0.05)} = 99 - 4 - (6 \cdot 2) + 1 = 84,
\end{equation}
and for the dataset with $\mathbf{B}_0=[0.10,0]$
\begin{equation}
N_{\text{seq}}^{(0.10)} = 59 - 4 - (6 \cdot 2) + 1 = 44.
\end{equation}

As a result, the total number of available sequences is : 
\begin{equation}
N_{\text{total}} = 84 + 44 = 128.
\end{equation}

These sequences are randomly split into training and validation sets using an $80\%$/$20\%$ ratio.
This results in
\begin{align}
N_{\text{train}} &= \lfloor 0.8 \times 128 \rfloor = 103 \quad \text{training sequences}, \\
N_{\text{val}}   &= 128 - 103 = 25 \quad \text{validation sequences}.
\end{align}

To ensure numerical stability during training and to enable consistent learning across datasets with different magnetic field strengths, the vorticity $\omega$ and current density $j$ fields are normalized prior to sequence extraction.
Global mean and standard deviation statistics are computed by aggregating all available training datasets, and each field is normalized as
\begin{equation}
\tilde{X} = \frac{X - \mu_X}{\sigma_X},
\end{equation}
where $X \in \{\omega, j\}$, and $\mu_X$ and $\sigma_X$ denote the corresponding global mean and standard deviation.
These statistics are computed once and reused for all datasets, including validation and test sets, ensuring a consistent normalization across different values of the imposed magnetic field $B_0$.

\subsection{Training Loss}

We use the Mean Squared Error (MSE) as the training objective.  
At each training step, the model predicts a block of $h$ future frames (in our case $h=2$).  
Training is performed autoregressively over $R$ rollouts (here $R=6$).  
For each rollout $R$, the model produces a sequence of predictions 
$\hat{y}_{h} \in \mathbb{R}^{H \times W}$ for $t = 1, \dots, h$, which are compared to the corresponding ground-truth frames $y_{t}$. Further details on the loss computation are given in the next section.

The MSE for a single rollout is defined as
\begin{equation}
\mathrm{MSE}_{r}(y, \hat{y}) = 
\frac{1}{h H W}
\sum_{t=1}^{h} 
\sum_{i=1}^{H}
\sum_{j=1}^{W}
\left( y_{t,i,j} - \hat{y}_{t,i,j} \right)^2 .
\label{eq:mse_rollout}
\end{equation}

This value corresponds to the pixel-wise squared error averaged jointly over all $h$ predicted frames.

The total training loss for one sequence is obtained by averaging the MSE over all autoregressive rollouts:
\begin{equation}
\mathcal{L} = 
\frac{1}{R} \sum_{r=1}^{R} \mathrm{MSE}_{r}.
\label{eq:training_loss}
\end{equation}

All models are trained for up to 500 epochs. Early stopping is employed to prevent overfitting by stopping training when the validation loss fails to improve. Furthermore, a learning rate scheduler is applied, reducing the learning rate when no further improvement in validation loss is observed in order to promote more stable convergence.

\subsection{Model size in terms of trainable parameters}

Table~\ref{tab:training_cost} indicates that both models remain parameter-efficient, with fewer than 5 million trainable parameters for high-resolution autoregressive MHD forecasting.
The Koopman Transformer uses $2.52\times 10^6$ trainable parameters and is trained in 114.61 minutes of GPU time, while the ConvLSTM-UNet uses $4.84\times 10^6$ parameters and is trained in 460.92 minutes on a single Tesla V100S-PCIE-32GB GPU.
Although ConvLSTM-UNet has more parameters ($\approx 1.92\times$ more) and longer training time, both models remain under 5 million parameters, with training durations between 2 and 8 hours on a single GPU, supporting their use in routine experimentation and deployment studies.

\begin{table*}[t]
\caption{Training cost comparison between the two forecasting models.}
\label{tab:training_cost}
\centering
\begin{tabular}{lccccc}
\hline
Model & Trainable parameters & Training GPU time & GPU \\
\hline
Koopman Transformer & 2,516,674 & 114.61 min & Tesla V100S-PCIE-32GB \\
ConvLSTM-UNet & 4,841,794 & 460.92 min & Tesla V100S-PCIE-32GB \\
\hline
\end{tabular}
\end{table*}

For high-resolution ($512\times512$) spatio-temporal forecasting of Kelvin-Helmholtz dynamics, these parameter counts are relatively low compared with many contemporary deep sequence models that rely on substantially larger capacities. 
This reduced model complexity is particularly relevant in our data-limited regime, where controlling model size helps to reduce over-parameterization risk while preserving sufficient expressive power to predict both vorticity and current density. Moreover, this trade-off is particularly important when evaluating robustness and generalization under limited training data.\\

\section{Performance Assessment of Data-Driven Models for 2D Incompressible Ideal MHD}
\label{sec:evaluation}

To objectively compare the forecasting performance of the KT-MHD and ConvLSTM-UNet models, we employ a common evaluation protocol combining statistical and physics-based consistency metrics to assess the strengths and limitations of each approach for data-driven MHD forecasting.
We consider our two-dimensional periodic domain $\mathcal{D}$ and define the spatial integral of a scalar field $f$: 

\begin{equation}
\langle f \rangle 
= \int_{\mathcal{D}} f(x,t)\, d\mathbf{x} \, ,
\end{equation}

with $d\mathbf{x}$ = $dx dy $.

\subsection{Conservation of 2D MHD Invariants}
\label{subsec:invariants}

In two-dimensional incompressible MHD certain global quantities are conserved in the ideal limit~\cite{frisch1975possibility,biskamp2003magnetohydrodynamic, davidson2017introduction} (i.e., in the absence of dissipation and resistivity). Two of the most important invariants are the total energy $E_{\rm tot}$ and the cross helicity $H_c$, which characterize the global dynamics of the velocity and magnetic fields.

The total energy is defined as
\begin{equation}
E_{\rm tot} = \frac{1}{2} \langle |\mathbf{u}|^2 \rangle + \frac{1}{2} \langle |\mathbf{B}|^2 \rangle,
\end{equation}

which represents the sum of kinetic and magnetic energies.

The cross helicity, which measures the degree of alignment between the velocity and magnetic fields, is defined as

\begin{equation}
H_c = \langle \mathbf{u} \cdot \mathbf{B} \rangle \, .
\end{equation}

In the ideal 2D MHD limit, both $E_{\rm tot}$ and $H_c$ are conserved over time, i.e.,
\begin{equation}
\frac{dE_{\rm tot}}{dt} = 0, \quad \frac{dH_c}{dt} = 0.
\end{equation}

In our numerical experiments, these invariants provide a physics-informed metric to assess the fidelity of autoregressive forecasts. Deviations from their initial values indicate the degree to which the model preserves fundamental MHD constraints over the predicted horizon.

\subsection{Energy Spectrum}
\label{subsec:energy_spectrum}

To assess whether the models preserve the correct physical distribution of energy across spatial scales, we compute the total energy spectra which is the sum of the kinetic and magnetic energy spectra, from the velocity and magnetic fields reconstructed from the predicted vorticity and current density fields.

The Fourier transform of the velocity field $\mathbf{u}(\mathbf{x})$ (and correspondingly for the magnetic field $\mathbf{B}(\mathbf{x}, t)$) is defined as:
\begin{equation}
\widehat{\mathbf{u}}(\mathbf{k}, t) = 
\int_{\mathcal{D}} \mathbf{u}(\mathbf{x}, t) \,
e^{-\iota \mathbf{k} \cdot \mathbf{x}} \, d\mathbf{x},
\end{equation}
where $\iota = \sqrt{-1}$ and $\mathbf{k} = (k_x, k_y)$ denotes the two-dimensional wavevector. 
The two-dimensional kinetic energy spectrum is then defined as:
\begin{equation}
E_{2D}(\mathbf{k}) = \frac{1}{2}
\left(
|\widehat{u}_x(\mathbf{k}, t)|^2 +
|\widehat{u}_y(\mathbf{k}, t)|^2
\right),
\end{equation}
and correspondingly the magnetic energy density as:
\begin{equation}
E^{B}_{2D}(\mathbf{k}) = \frac{1}{2}
\left(
|\widehat{B}_x(\mathbf{k}, t)|^2 +
|\widehat{B}_y(\mathbf{k}, t)|^2
\right).
\end{equation}

A radial average in Fourier space yields the one-dimensional isotropic energy spectra:
\begin{equation}
E_u(k) = \sum_{k - \frac{1}{2} \le |\mathbf{k}| < k + \frac{1}{2}}
E_{2D}(\mathbf{k}),
\end{equation}
and for $E_B(k)$ accordingly.

The total one-dimensional energy spectrum is then defined as the sum of kinetic and magnetic contributions:
\begin{equation}
E(k) = E_u(k) + E_B(k).
\end{equation}

The energy spectra quantify the distribution of kinetic and magnetic energy across wavenumbers and permit to identify powerlaw scaling and wavenumbers where dissipative effects become dominant.

To facilitate the analysis of scaling properties, we plot compensated spectra $k^{3/2} E(k)$. 
In MHD turbulence, the inertial range energy cascade is expected to follow the scaling law $E(k) \propto k^{-3/2}$, see Ref.~\cite{biskamp2003magnetohydrodynamic}. 
Multiplication by $k^{3/2}$ thus yields an approximately constant value within the inertial range. 

Comparing predicted and reference spectra therefore enables an evaluation of whether the learned models correctly reproduce the multiscale energy transfer mechanisms inherent to Kelvin-Helmholtz-driven MHD turbulence. In particular, a model is characterized as overly dissipative when $E_{\mathrm{pred}}(k) < E_{\mathrm{true}}(k)$ at large wavenumbers, indicating excessive damping of small-scale structures.

\subsection{Enstrophy and magnetic enstrophy}
\label{subsec:Enstrophy_magnetic_enstrophy}

The enstrophy and magnetic enstrophy are defined as
\begin{equation}
\Omega_\omega = \frac{1}{2} \langle \omega^2 \rangle, 
\quad 
\Omega_j = \frac{1}{2} \langle j^2 \rangle \, .
\end{equation}

The enstrophy $\Omega_\omega$ quantifies the mean-square vorticity and thus measures the intensity of 
rotational motion and velocity gradients in the flow. In two-dimensional turbulence, it is directly associated with the activity of fine-scale structures, since vorticity concentrates at progressively smaller scales during the enstrophy cascade.

Similarly, the magnetic enstrophy $\Omega_j$ measures the mean-square current density and provides a diagnostic of small-scale magnetic gradients, including the formation of current sheets and regions of strong magnetic shear.

Both quantities are therefore particularly sensitive to the development of small-scale structures in MHD turbulence. 

\subsection{Evaluation Metrics}
\label{subsec:metrics}

To quantitatively assess the accuracy and physical reliability of the autoregressive forecasts, we employ complementary error metrics that capture both pointwise numerical deviations and structural fidelity.

\paragraph{Mean Squared Error (MSE).}

The Mean Squared Error quantifies the average pixel-wise discrepancy between the predicted field  $\hat{X}$ and the reference field $X$. 
For a single 2D frame of size $H\times W$, the MSE is defined as
\begin{equation}
\mathrm{MSE}(X,\hat{X}) = 
\frac{1}{HW}
\sum_{i=1}^{H}\sum_{j=1}^{W}
\left( X_{ij} - \hat{X}_{ij} \right)^2 .
\end{equation}

In the autoregressive setting, predictions are generated over multiple timesteps $t$ and test sequences $n$. The timestep- and sequence-dependent MSE is therefore
\begin{equation}
\mathrm{MSE}_{n,t} =
\frac{1}{HW}
\sum_{i=1}^{H}\sum_{j=1}^{W}
\left( X^{(n,t)}_{ij} - \hat{X}^{(n,t)}_{ij} \right)^2 .
\end{equation}

The global MSE reported in our analysis corresponds to the average over all test sequences and prediction timesteps:
\begin{equation}
\mathrm{MSE}_{\mathrm{global}} =
\frac{1}{N_{\mathrm{seq}} T}
\sum_{n=1}^{N_{\mathrm{seq}}}
\sum_{t=1}^{T}
\mathrm{MSE}_{n,t}.
\end{equation}

Lower MSE values indicate better reconstruction accuracy. Due to its quadratic nature, the MSE strongly penalizes large-amplitude local errors and is therefore sensitive to high-frequency discrepancies.\\

The global Root Mean Squared Error is defined as the square root of the global MSE,
\begin{equation}
\mathrm{RMSE}_{\mathrm{global}} =
\sqrt{\mathrm{MSE}_{\mathrm{global}}},
\end{equation}
and retains the same physical units as the original field, facilitating physical interpretability.\\

\paragraph{Mean Absolute Error (MAE).}

To provide a complementary measure less sensitive to localized large deviations, we also consider the Mean Absolute Error. For consistency with the above notation, the global MAE is defined as
\begin{equation}
\mathrm{MAE}_{\mathrm{global}} =
\frac{1}{N_{\mathrm{seq}} T}
\sum_{n=1}^{N_{\mathrm{seq}}}
\sum_{t=1}^{T}
\left(
\frac{1}{HW}
\sum_{i=1}^{H}\sum_{j=1}^{W}
\left| X^{(n,t)}_{ij} - \hat{X}^{(n,t)}_{ij} \right|
\right).
\end{equation}

\paragraph{Peak Signal-to-Noise Ratio (PSNR).}
PSNR quantifies the fidelity of a reconstructed field relative to its dynamic range. 
For a given frame, let $X$ denote the reference field (ground truth) and $\hat{X}$ its prediction. 
We define the dynamic range as
\begin{equation}
L = \max(X) - \min(X).
\end{equation}
The PSNR is then computed as
\begin{equation}
\mathrm{PSNR}(X,\hat{X})
= 10 \log_{10} \left( \frac{L^2}{\mathrm{MSE}(X,\hat{X})} \right),
\end{equation}
 
A higher PSNR indicates better global agreement and lower noise level. While originally introduced for digital imaging, PSNR is effective for quantifying error magnitude in smoothly varying physical fields such as vorticity or current density.\\

\paragraph{Structural Similarity Index (SSIM).}
SSIM measures perceptual similarity by comparing local spatial statistics between the reference field $X$ and the prediction $\hat{X}$. 
For two local patches $x$ and $\hat{x}$, the SSIM is defined as
\begin{equation}
\mathrm{SSIM}(x,\hat{x}) =
\frac{ (2\mu_x \mu_{\hat{x}} + C_1)(2\sigma_{x\hat{x}} + C_2) }
     { (\mu_x^2 + \mu_{\hat{x}}^2 + C_1)(\sigma_x^2 + \sigma_{\hat{x}}^2 + C_2) },
\end{equation}
where $\mu_x$ and $\mu_{\hat{x}}$ denote local means, 
$\sigma_x^2$ and $\sigma_{\hat{x}}^2$ the local variances, 
$\sigma_{x\hat{x}}$ the local covariance, 
and $C_1$, $C_2$ small constants for numerical stability~\cite{wang2004image}.
In this current work, the $C_1$ and $C_2$ are defined as
\[
C_1 = (K_1 L)^2, \quad C_2 = (K_2 L)^2,
\]
where $L$ denotes the dynamic range of the field (i.e., $L = \max - \min$), and $K_1 = 0.01$, $K_2 = 0.03$. 
  
The final SSIM score is obtained by averaging these local values over all spatial positions.  
This metric is particularly relevant for MHD flows because it captures the preservation of coherent structures such as vortex sheets, current layers, and filaments that govern nonlinear interactions and energy transfer.\\

Together, these metrics provide a comprehensive and complementary assessment of forecast quality.
By using all these metrics, the comparison between KT-MHD and ConvLSTM-UNet reflects not only pixel level reconstruction performance but also the model’s ability to reproduce the physically relevant patterns that govern Kelvin-–Helmholtz dynamics.

\section{Quantifying Structural and Spectral Accuracy in 2D MHD Flow Forecasts}
\label{sec:results}

This section reports the predictive performance of the two deep learning architectures introduced in Section~\ref{sec:models}: the Koopman Transformer for MHD (KT-MHD) and the ConvLSTM U-Net. We present the input sequence and target frames, show examples of autoregressive predictions produced by both models, spectral diagnostics for representative predicted frames, a quantitative comparison (MSE, SSIM, PSNR) and computational cost, and an interpretation and discussion of the results.\\

To assess the generalization and robustness of the proposed models, predictions are evaluated for simulations with magnetic field strengths $B_0 = [0.08,0]$ and $B_0 = [0.12,0]$. These configurations respectively correspond to interpolation and extrapolation cases with respect to the training set, and allow us to examine whether the models preserve physically meaningful behavior when applied to unseen magnetic regimes.
To avoid the presentation of too many results and figures, the results obtained for $\mathbf{B}_0 = [0.08, 0]$ are reported separately in the Appendix.

\begin{figure*}[ht]
    \centering

    % -- Ligne 1 --
    \begin{subfigure}{0.95\linewidth}
        \centering
        \includegraphics[width=0.9\linewidth]{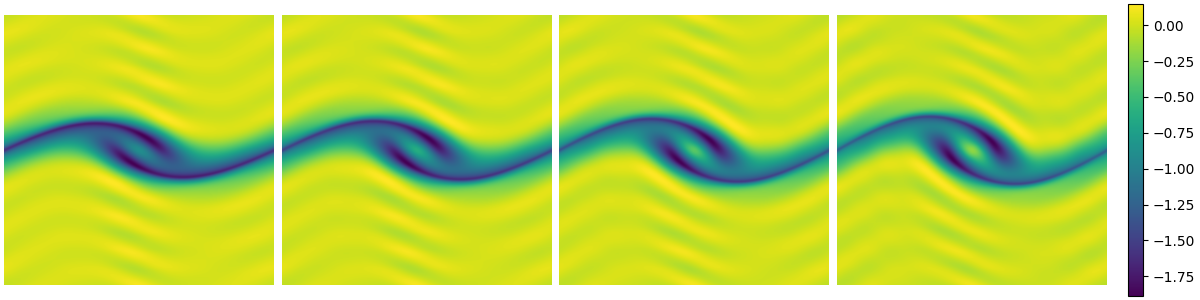}
        \caption{Input sequence for the vorticity field $\omega$}
    \end{subfigure}

    \vspace{1em}

    % -- Ligne 2 --
    \begin{subfigure}{0.95\linewidth}
        \centering
        \includegraphics[width=0.9\linewidth]{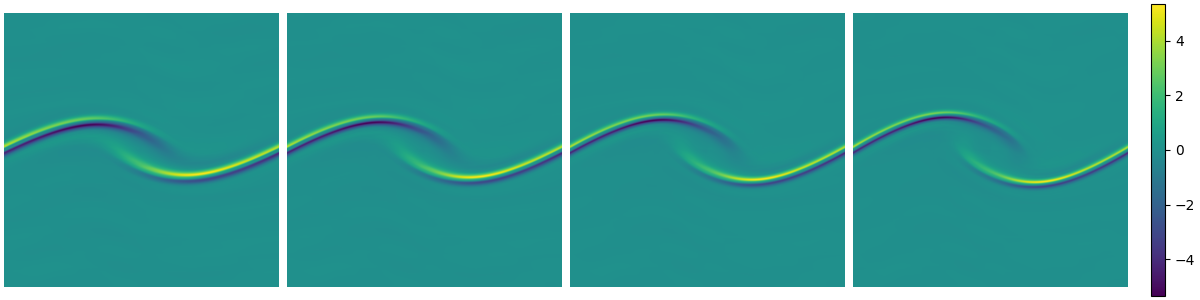}
        \caption{Input sequence for the current density field $j$}
    \end{subfigure}

    \caption{Input sequences of four past frames for  vorticity $\omega$ and current density $j$, with an initial magnetic field $\mathbf{B}_0 = [0.12, 0]$. The inputs correspond to consecutive time steps $t = 20, 20.5, 21$ and $21.5$.
}
    \label{fig:input_sequence_wj_B0_0_12}
\end{figure*}

\begin{figure*}
\includegraphics[width=1\linewidth]{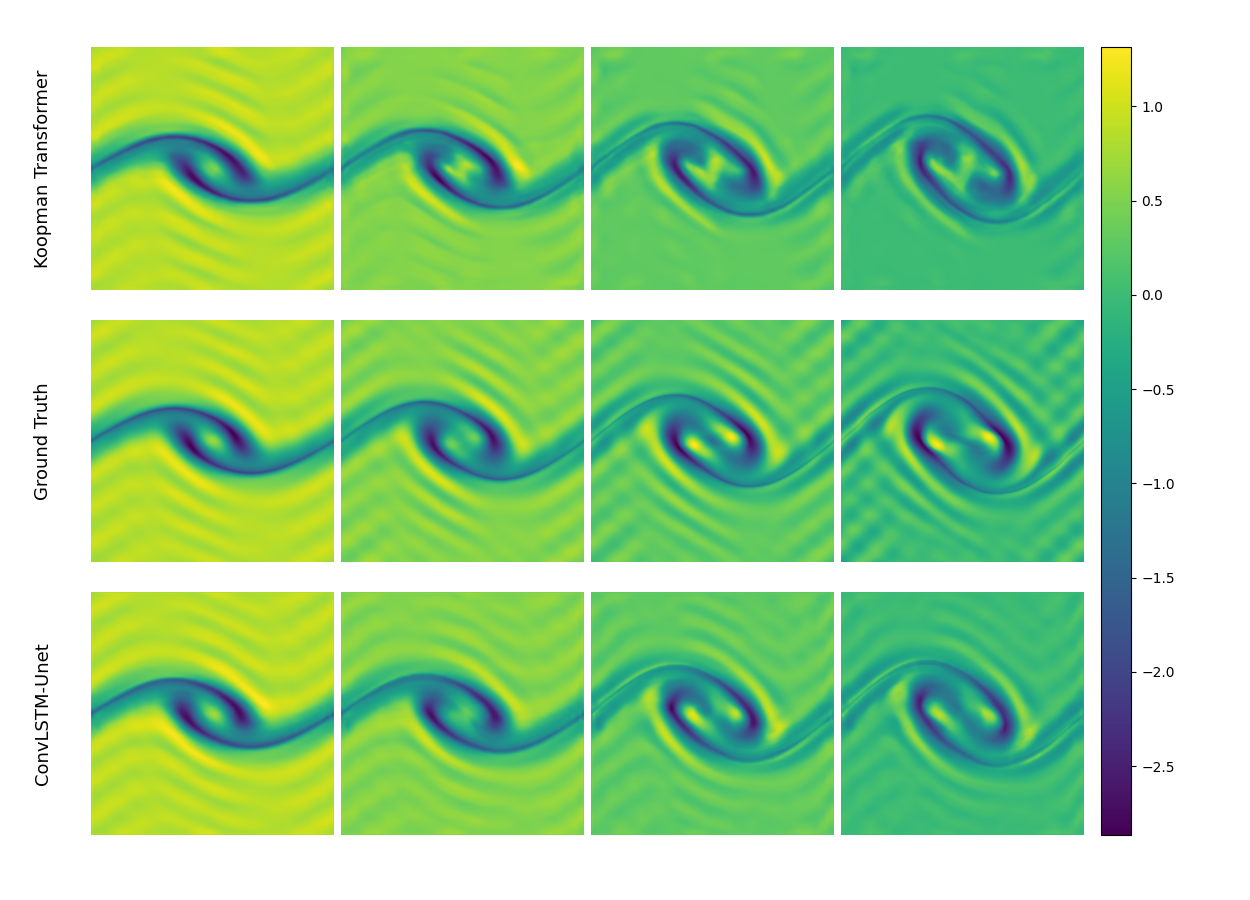}
\caption{\label{fig:vorticity_prediction_KT_LSTM_Unet_B0_0_12} Autoregressive forecasting of the vorticity field $\omega$ for $B_0=[0.12,0]$. 
    From top to bottom: KT-MHD Transformer prediction, Ground Truth (DNS reference), and ConvLSTM U-Net prediction. For visualization purposes, only every third frame is shown, corresponding to time steps $t = 22, 22.5, 23, \dots , 27.5$.
}
\end{figure*}

\begin{figure*}
\includegraphics[width=1\linewidth]{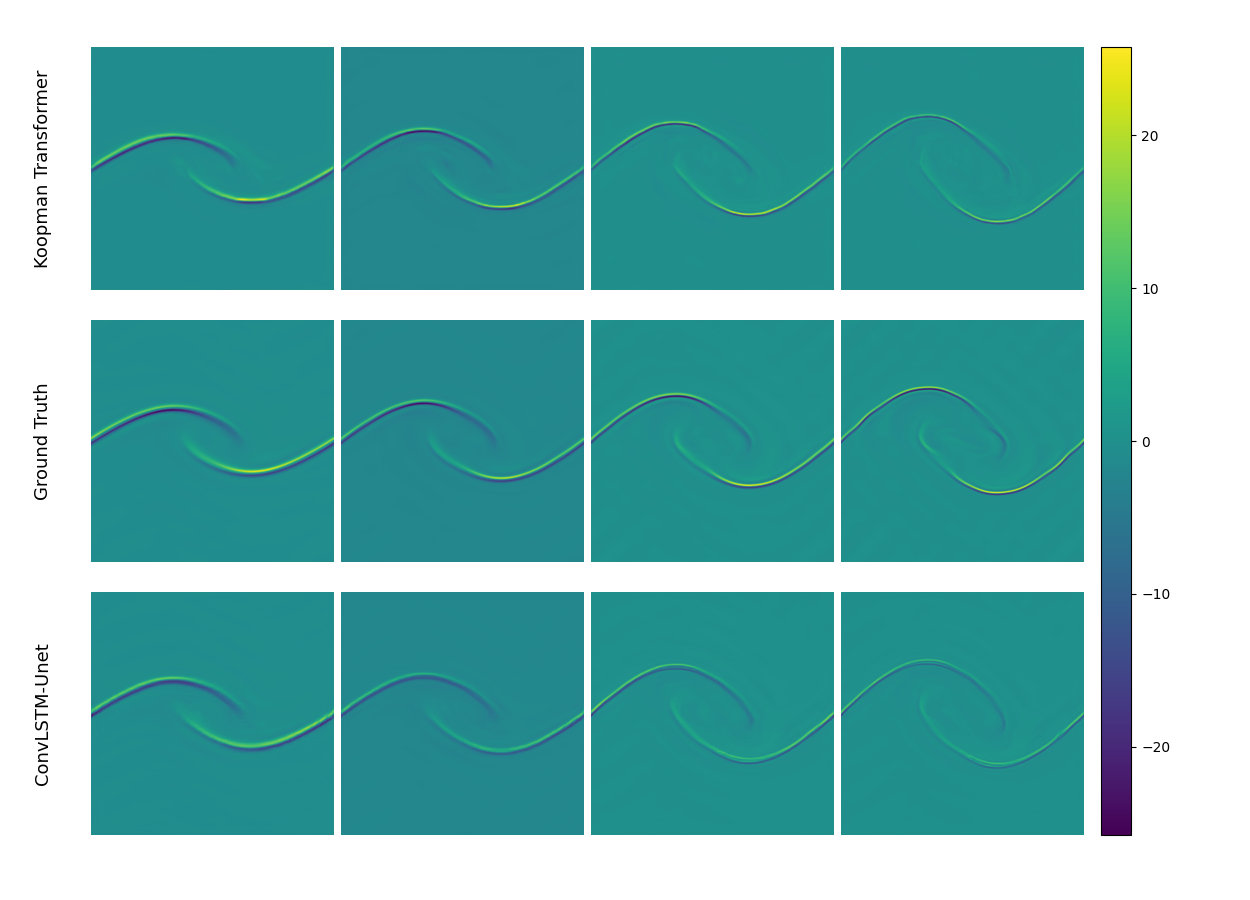}
\caption{\label{fig:current_prediction_KT_LSTM_Unet_B0_0_12} Autoregressive forecasting of the current density field $j$ for $B_0=[0.12,0]$. 
    From top to bottom: KT-MHD Transformer prediction, Ground Truth (DNS reference), and ConvLSTM U-Net prediction. For visualization purposes, only every third frame is shown, corresponding to time steps $t = 22, 22.5, 23, \dots , 27.5$.}
\end{figure*}

\begin{table}[ht]
\centering
\caption{Performance metrics for the Koopman Transformer model on the predicted frames for the vorticity field shown in figure~\ref{fig:vorticity_prediction_KT_LSTM_Unet_B0_0_12}. } 
\label{tab:koopman_vorticity_metrics}
\renewcommand{\arraystretch}{1.7}
\resizebox{0.9\linewidth}{!}{
\begin{tabular}{c c c c c c}
\hline
Timestep t & MSE & RMSE & MAE & SSIM & PSNR \\
\hline
22 & 0.0009 & 0.030 & 0.019 & 0.906 & 36.91 \\
23.5 & 0.006 & 0.079 & 0.055 & 0.662 & 30.58 \\
25 & 0.023 & 0.152 & 0.099 & 0.505 & 25.54 \\
26.5 & 0.054  & 0.233  & 0.163  & 0.384  & 23.30  \\
\hline
\end{tabular}}
\end{table}

\begin{table}[ht]
\centering
\caption{Performance metrics for the ConvLSTM-UNet model on the predicted frames for the vorticity field shown in figure~\ref{fig:vorticity_prediction_KT_LSTM_Unet_B0_0_12}.}
\label{tab:lstmunet_vorticity_metrics}
\renewcommand{\arraystretch}{1.7}
\resizebox{0.9\linewidth}{!}{
\begin{tabular}{c c c c c c}
\hline
Timestep t & MSE & RMSE & MAE & SSIM & PSNR \\
\hline
22 & 0.0020 & 0.045 & 0.025 & 0.923 & 33.42 \\
23.5 & 0.0043 & 0.066 & 0.044 & 0.762 & 31.15 \\
25 & 0.011 & 0.108 & 0.078 & 0.661 & 29.22 \\
26.5 & 0.037   & 0.194  & 0.135  & 0.524  & 25.08  \\
\hline
\end{tabular}}
\end{table}

\begin{table}[ht]
\centering
\caption{Performance metrics for the Koopman Transformer model on the predicted frames for the current density field shown in figure~\ref{fig:current_prediction_KT_LSTM_Unet_B0_0_12}.} 
\label{tab:koopman_current_density_metrics}
\renewcommand{\arraystretch}{1.7}
\resizebox{0.9\linewidth}{!}{
\begin{tabular}{c c c c c c}
\hline
Timestep t & MSE & RMSE & MAE & SSIM & PSNR \\
\hline
22 & 0.046 & 0.215 & 0.079 & 0.882 & 35.55 \\
23.5 & 0.148 & 0.385 & 0.126 & 0.895 & 35.81 \\
25 & 0.636 & 0.798 & 0.219 & 0.861 & 31.96 \\
26.5 & 2.108   & 1.452  & 0.407  & 0.773  & 27.82  \\
\hline
\end{tabular}}
\end{table}

% Tableau suivant pour LSTM-UNet
\begin{table}[ht]
\centering
\caption{Performance metrics for the ConvLSTM-UNet model on the predicted frames for the current density field shown in figure~\ref{fig:current_prediction_KT_LSTM_Unet_B0_0_12}.} 
\label{tab:lstmunet_current_density_metrics}
\renewcommand{\arraystretch}{1.7}
\resizebox{0.9\linewidth}{!}{
\begin{tabular}{c c c c c c}
\hline
Timestep t & MSE & RMSE & MAE & SSIM & PSNR \\
\hline
22 & 0.106 & 0.327 & 0.088 & 0.907 & 30.85 \\
23.5 & 0.179 & 0.424 & 0.116 & 0.893 & 30.88 \\
25 & 1.324 & 1.151 & 0.279 & 0.865 & 27.09 \\
26.5 & 2.411   & 1.553  & 0.388  & 0.832  & 27.07  \\
\hline
\end{tabular}}
\end{table}

\subsubsection{Qualitative analysis of autoregressive predictions}
\label{subsec:Qualitative analysis of autoregressive predictions modified3}

Figure~\ref{fig:input_sequence_wj_B0_0_12} shows the input sequences for the vorticity $\omega$ and current density $j$ for $B_0 = [0.12,0]$. The first row corresponds to vorticity $\omega$, characterized by a coherent central shear layer with smooth large-scale evolution, while the second row corresponds to current density $j$, which exhibits finer and more irregular small-scale structures. These inputs represent complementary MHD dynamics: $\omega$ captures large-scale flow organization, whereas $j$ reflects magnetic gradients and small-scale current structures. Using the same temporal window for both fields provides the model with consistent coupling information for autoregressive forecasting.

Figures~\ref{fig:vorticity_prediction_KT_LSTM_Unet_B0_0_12} and~\ref{fig:current_prediction_KT_LSTM_Unet_B0_0_12} present autoregressive predictions for both variables and both models. Although the rollout spans 12 forecasted frames (horizon $=2$, rollout $=6$), only one out of every three frames is shown for clarity.

Quantitative performance is reported in Tables~\ref{tab:koopman_vorticity_metrics}--\ref{tab:lstmunet_current_density_metrics}. These results indicate that the ConvLSTM-UNet achieves better accuracy for the vorticity field $\omega$, with lower MSE, RMSE, and MAE values, along with higher SSIM and PSNR scores across all timesteps, with a PSNR of 29.22 at timestep t=25 compared to 25.54 for the Koopman Transformer, as reported in tables~\ref{tab:koopman_vorticity_metrics} and \ref{tab:lstmunet_vorticity_metrics}, while the Koopman Transformer performs better for the current density $j$, particularly at longer horizons.

These trends are consistent with the qualitative observations, both models correctly reproduce the large-scale evolution of the vorticity field, including the deformation of the central shear layer and the global phase of the instability. The ConvLSTM-UNet preserves fine-scale structures more effectively, maintaining sharper gradients and small-scale features over time. This is reflected quantitatively by its lower prediction errors compared to the Koopman Transformer (e.g., MSE $=0.011$ vs $0.023$ at timestep t=25, Tables~\ref{tab:koopman_vorticity_metrics}--\ref{tab:lstmunet_vorticity_metrics}). 

In contrast, the KT-MHD Transformer produces smoother fields, with a faster loss of small-scale details but improved large-scale consistency. This progressive smoothing is reflected in the quantitative metrics reported in tables~\ref{tab:koopman_current_density_metrics} and~\ref{tab:lstmunet_current_density_metrics}, where the PSNR decreases from 35.55 dB at the first prediction step t=22 to 27.82 dB at later timesteps. Such degradation is expected in autoregressive forecasting due to error accumulation. Nevertheless, the Transformer maintains competitive performance, particularly for the current density field, where it remains more stable than the ConvLSTM-UNet at longer prediction horizons.

The differences are more pronounced for the current density $j$, which is inherently more sensitive to small-scale variations. The ConvLSTM-UNet predictions progressively lose contrast and exhibit thickening of current layers, indicating diffusion of fine magnetic structures. This behavior is consistent with its higher error levels at later timesteps (e.g., MSE $=2.411$ at timestep t=26.5). The Koopman Transformer, on the other hand, maintains sharper current sheets and more coherent structures over the autoregressive rollout, which is supported by its lower MSE values (e.g., $2.108$ vs $2.411$ at timestep t=26.5 (Tables~\ref{tab:koopman_current_density_metrics}--\ref{tab:lstmunet_current_density_metrics}).

Overall, both models well capture the nonlinear evolution of the Kelvin-Helmholtz instability and produce stable multi-step predictions. However, they exhibit complementary strengths: the ConvLSTM-UNet is more effective at preserving localized, high-gradient structures in the vorticity field, while the KT-MHD Transformer better captures the global organization and temporal evolution of magnetic structures. This behavior reflects the different inductive biases of the architectures, suggesting that combining local and global modeling strategies could further improve multi-variable MHD forecasting.

\subsubsection{Error accumulation and enstrophy trends in 2D MHD forecasts}

In this section, we examine the MSE and enstrophy evolution along autoregressive predictions for both the Koopman transformer and the ConvLSTM-UNet models, in order to provide insight into their respective capabilities for long-term forecasting of MHD dynamics.\\

\begin{figure*}[ht]
    \centering

    \begin{subfigure}{0.49\linewidth}
        \centering
        \includegraphics[width=\linewidth]{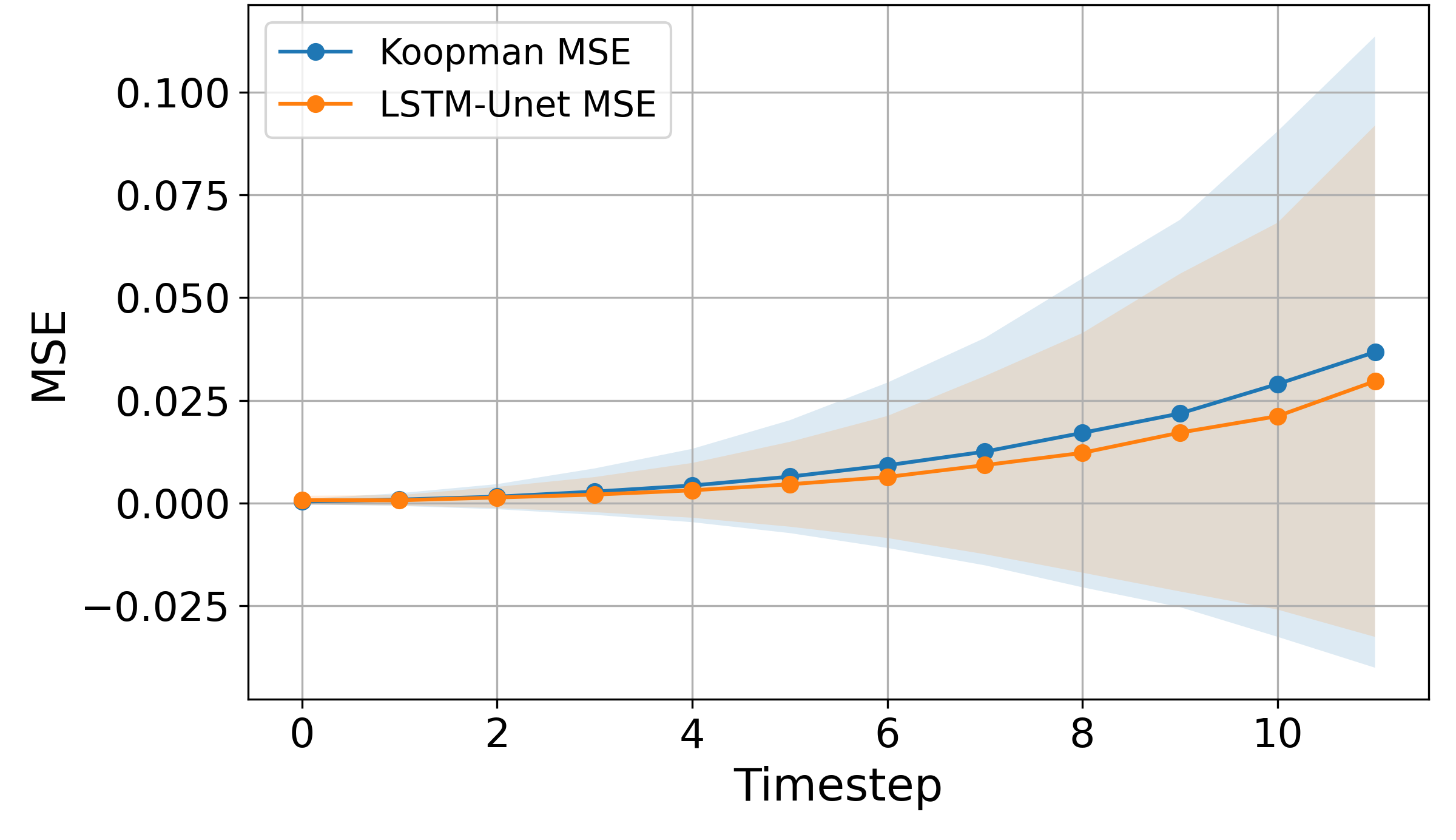}
        \caption{Evolution of MSE over time of the predicted vorticity field for Koopman and ConvLSTM-UNet. }
    \end{subfigure}
    \hfill
    \begin{subfigure}{0.49\linewidth}
        \centering
        \includegraphics[width=\linewidth]{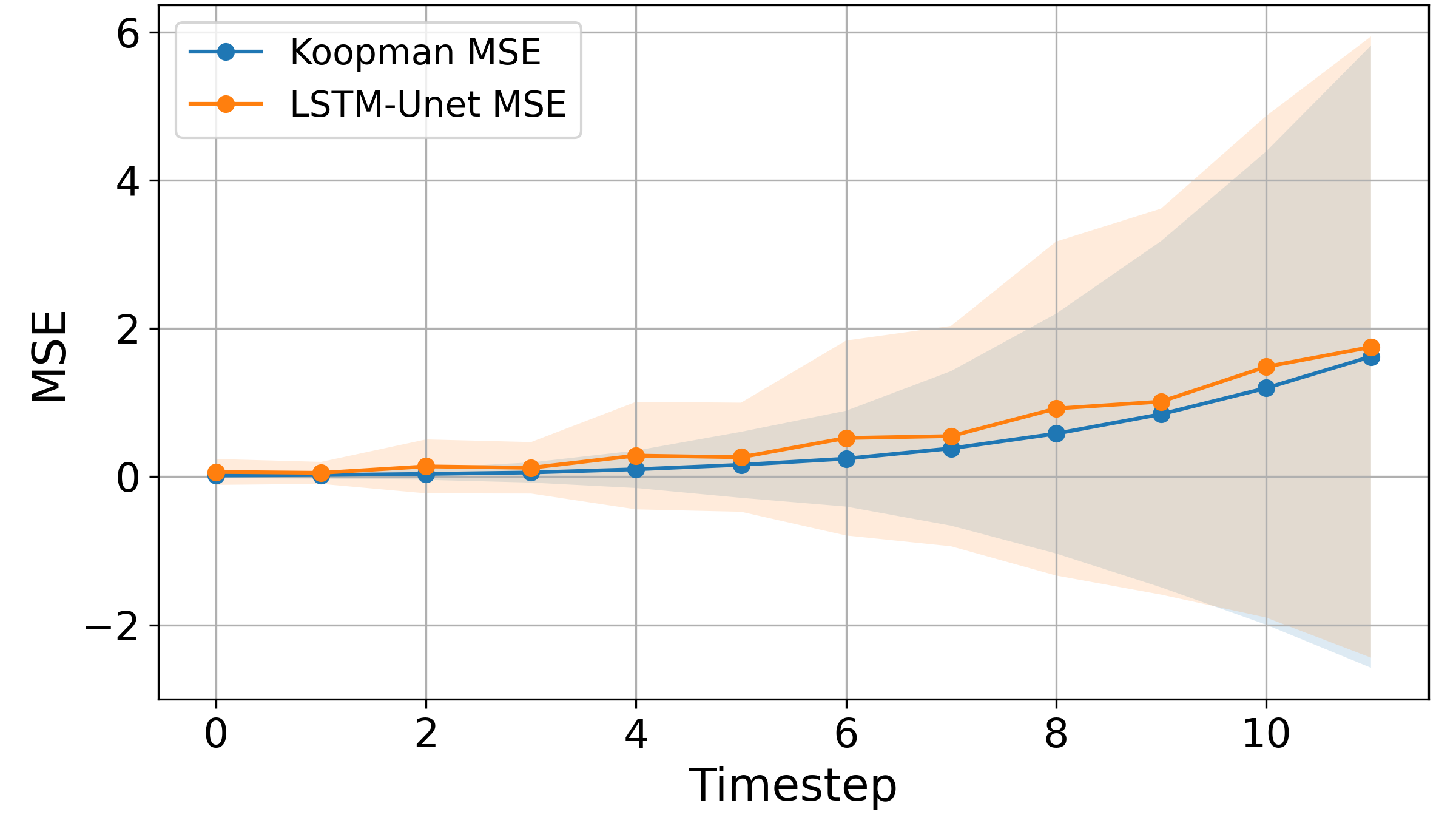}
        \caption{Evolution of MSE over time of the predicted current density field for Koopman and ConvLSTM-UNet.}
    \end{subfigure}

    \caption{Mean (solid lines) and standard deviation (shaded $\pm1\sigma$) of MSE at each rollout time step for $\omega$ and $j$, with an initial magnetic field $B_0 = [0.12,0]$.}
    \label{fig:Mean_MSE_at_each_rollout_time_step_KT_wj_B0_0_12}
\end{figure*}

\paragraph{Propagation of prediction errors}
Figure~\ref{fig:Mean_MSE_at_each_rollout_time_step_KT_wj_B0_0_12} shows the evolution of the MSE as a function of the autoregressive timestep for both vorticity $\omega$ and current density $j$, averaged over all test sequences for the Koopman Transformer and ConvLSTM-UNet models. The error is computed pixel-wise on the 2D spatial grid and then averaged across sequences (Section~\ref{sec:evaluation}), while the shaded area represents the mean value plus/minus one standard deviation, quantifying the variability of the prediction..

For both variables $\omega$ and $j$, the MSE increases monotonically with the prediction horizon for both models, which is a characteristic behavior of autoregressive forecasting in chaotic dynamical systems~\cite{vlachas2018data}. This trend reflects the progressive accumulation of small local errors as model predictions are recursively fed back as inputs. The growth of the standard deviation at later timesteps indicates an increasing dispersion of forecast quality among different sequences, consistent with the sensitivity to initial conditions inherent to Kelvin-Helmholtz driven MHD.

A clearer distinction between the two variables is observed for the current density $j$, which exhibits a faster error growth and larger variance compared to the vorticity field. This is explained by its richer small-scale content and stronger dependence on spatial gradients of the magnetic field. As a consequence, even mild numerical diffusion or spatial smoothing in the prediction leads to a stronger degradation of $j$, which is typically manifested by an underestimation of magnetic enstrophy and an attenuation of high-wavenumber energy content, as further discussed in Subsection~\ref{subsec:energy_spectrum_analysis}.

A comparison between the two models shows that the ConvLSTM-UNet yields slightly lower MSE for the vorticity field (e.g., $0.011$ vs $0.023$ at timestep 3, Table~\ref{tab:koopman_vorticity_metrics}--\ref{tab:lstmunet_vorticity_metrics}), whereas the Koopman Transformer performs better for the current density at later timesteps (e.g., $2.108$ vs $2.411$ at timestep 4, Table~\ref{tab:koopman_current_density_metrics}--\ref{tab:lstmunet_current_density_metrics}). These differences remain moderate, indicating comparable pointwise accuracy, as also reflected by SSIM and PSNR values reported in Tables~\ref{tab:koopman_vorticity_metrics}--\ref{tab:lstmunet_current_density_metrics}.\\

\paragraph{Enstrophy and magnetic enstrophy}

\begin{figure}
\includegraphics[width=1\linewidth]{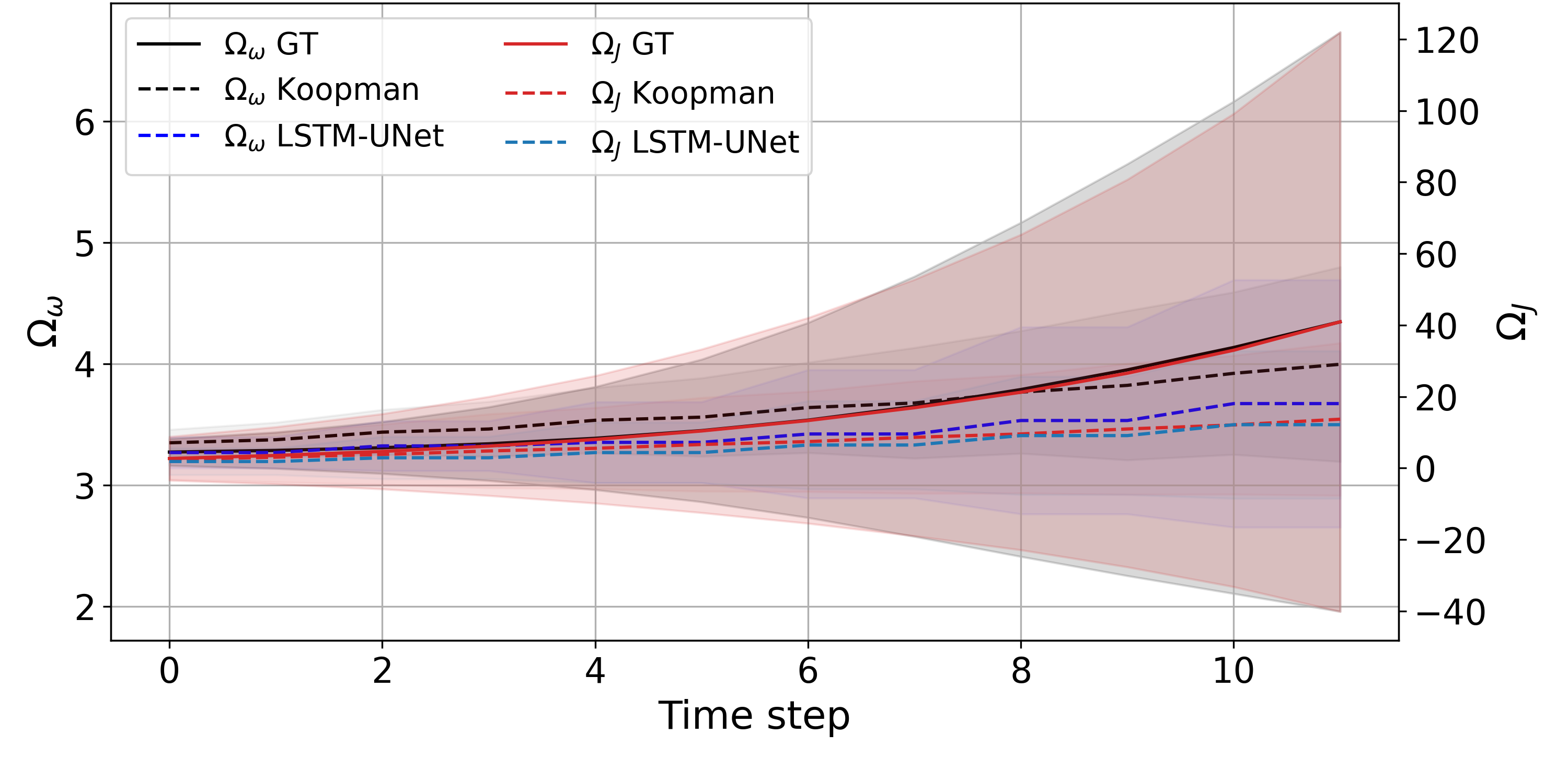}

\caption{\label{fig:KT_LSTM_Unet_enstrophy_wj} Evolution of the mean of averaged enstrophy and magnetic enstrophy for the ground truth (GT), Koopman and ConvLSTM-UNet. The solid line shows the mean across (number of samples/realizations), while the shaded region (error bars) represents $\pm1$ standard deviation.}

\end{figure}

We now analyze the temporal evolution of enstrophy $\Omega_\omega$ and magnetic enstrophy $\Omega_j$ computed from the autoregressively generated fields (Figure~\ref{fig:KT_LSTM_Unet_enstrophy_wj}). The shaded regions represent one standard deviation, while the reference DNS curves exhibit the expected Kelvin-Helmholtz evolution, characterized by an initial growth phase followed by nonlinear saturation.

Both models reproduce the global temporal dynamics of $\Omega_\omega$ and $\Omega_j$, with small deviations at later times, this small underestimation is expected, as autoregressive models tend to smooth fine-scale turbulent structures over time.\\

For the vorticity enstrophy $\Omega_\omega$, the ConvLSTM-UNet provides sharper agreement in the early nonlinear regime, consistent with its better preservation of fine-scale velocity structures observed in spatial fields and reflected in higher SSIM values (Table~\ref{tab:lstmunet_vorticity_metrics}). In contrast, the Koopman Transformer shows a slightly smoother evolution, with a mild overestimation during the growth phase followed by a small underestimation at later times, while remaining closer to the DNS curve in terms of global trend.

For the magnetic enstrophy $\Omega_j$, the Koopman Transformer achieves better overall agreement with the reference, which can be seen with its lower MSE values and improved reconstruction of current density structures (Table~\ref{tab:koopman_current_density_metrics}). The ConvLSTM-UNet, while capturing the correct trend, exhibits a stronger underestimation at later times, as indicated by the diffusion of small-scale magnetic structures observed in the spatial predictions in figure~\ref{fig:current_prediction_KT_LSTM_Unet_B0_0_12}.

Overall, these results highlight a fundamental distinction between local and global diagnostics. While spatial metrics and spectral indicators emphasize the reconstruction of small-scale features, enstrophy provides a global quadratic measure that is highly sensitive to amplitude errors. Consequently, a model that better preserves fine-scale structures does not necessarily achieve the closest agreement in integrated quantities such as enstrophy.

Despite these differences, both models correctly capture the dominant physical mechanisms of the Kelvin-Helmholtz instability, including growth and saturation phases, indicating robust long-term physical consistency.

\subsubsection{Tracking 2D MHD Invariants}

To complement the spectral analysis, we monitor the total energy $E_{\rm tot}$ and cross helicity $H_c$ during the autoregressive rollouts.\\

\paragraph{Invariant-Based Error Metrics}
\label{subsec:invariant_metrics}

While the temporal evolution of $E_{\rm tot}$ and $H_c$ provides qualitative insight into the physical fidelity of the forecasts, a quantitative comparison between models requires dedicated invariant-based error metrics. 
We therefore introduce complementary indicators that measure relative deviations, long-term drift, and normalized reconstruction errors of the predicted invariants.

\subparagraph{Mean Relative Error.}
The mean relative error quantifies the deviation as a proportion of the signal amplitude. 
It is particularly useful when comparing quantities with different orders of magnitude, such as the total energy $E_{\rm tot}$, the cross helicity $H_c$.
For a given invariant $X \in \{E_{\rm tot}, H_c\}$, the relative error for sequence $i$ at time $t$ is defined as
\begin{equation}
\mathrm{RelErr}_{i,t} =
\frac{\left| X^{\mathrm{pred}}_{i,t} - X^{\mathrm{gt}}_{i,t} \right|}
{\left| X^{\mathrm{gt}}_{i,t} \right| + \epsilon},
\end{equation}
where $\epsilon$ is a small constant introduced to avoid division by zero.
The reported mean relative error corresponds to an average over all test sequences and timesteps,
\begin{equation}
\mathrm{MeanRelErr} =
\left\langle \mathrm{RelErr}_{i,t} \right\rangle_{i,t}.
\end{equation}

\subparagraph{Mean Drift.}

To quantify the long-term conservation properties of the models, we define the drift of an invariant $X$ over a rollout of length $T$ for each test sequence $i$ as
\begin{equation}
\mathrm{Drift}_i =
\frac{X^{\mathrm{pred}}_{i,T-1} - X^{\mathrm{pred}}_{i,0}}
{X^{\mathrm{pred}}_{i,0} + \epsilon}.
\end{equation}
The mean drift is then obtained by averaging over all test sequences,
\begin{equation}
\mathrm{MeanDrift} =
\left\langle \mathrm{Drift}_i \right\rangle_i.
\end{equation}

\vspace{-0.5em}

The mean drift is especially relevant for physical invariants such as energy and helicity, where accumulated drift over time represents a key failure mode of autoregressive models. Its interpretation is intuitive: it answers whether the model preserves the invariant over long time horizons.

\begin{figure} 
    \centering

    % -- Ligne 1 --
    \begin{subfigure}{0.99\linewidth}
        \centering
        \includegraphics[width=1.02\linewidth]{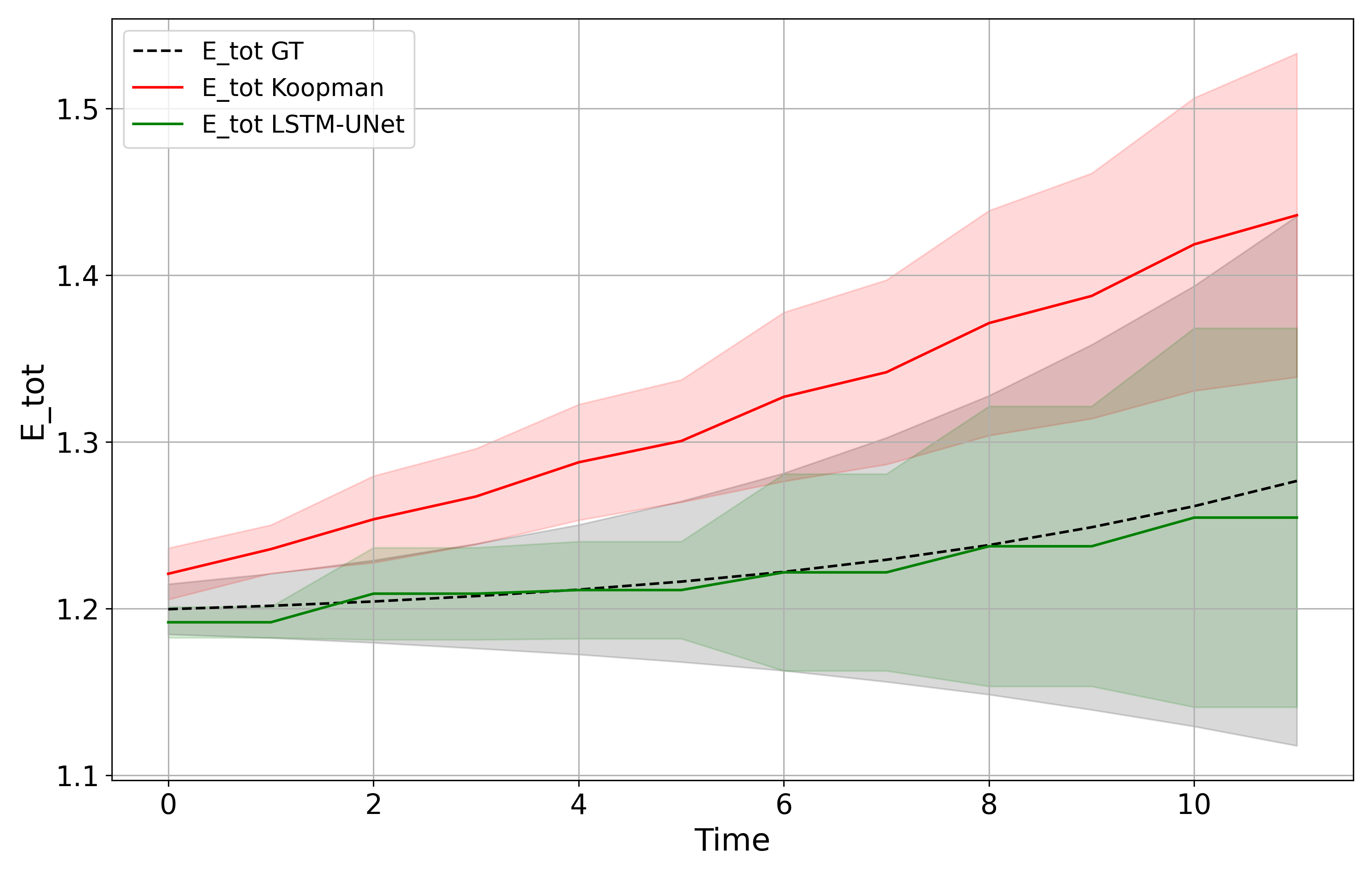}
        \caption{Evolution of the total energy ($E_{tot}$) over time for both models against the ground truth.}
    \end{subfigure}

    \vspace{1em}

    % -- Ligne 2 --
    \begin{subfigure}{0.99\linewidth}
        \centering
        \includegraphics[width=1.02\linewidth]{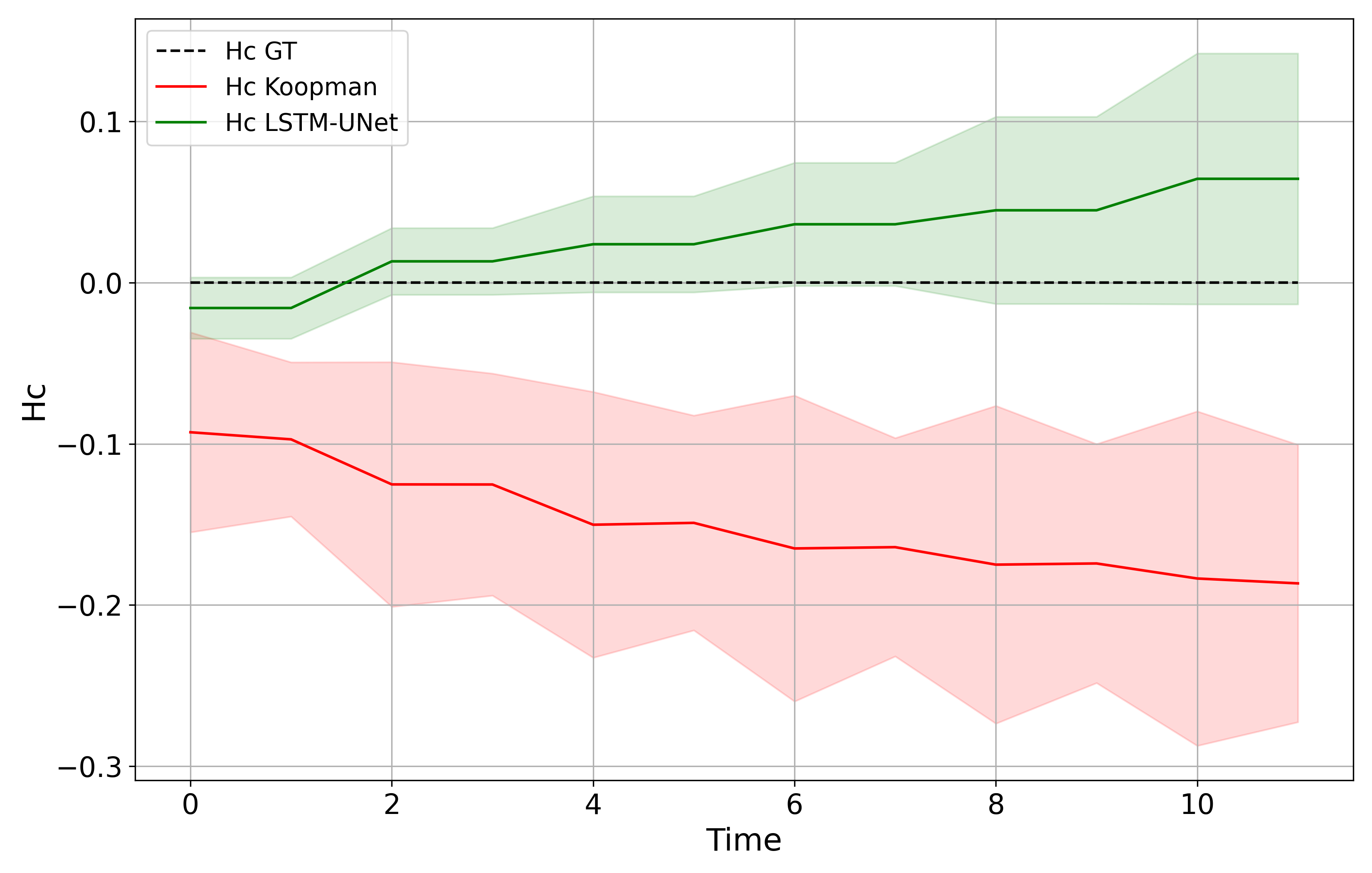}
        \caption{Evolution of the cross helicity ($H_c$) over time for both models against the ground truth.}
    \end{subfigure}

    \caption{Evolution of the mean of MHD invariant, for an initial magnetic field $B_0 =[0.12,0]$. The shaded region represents $\pm1$ standard deviation.}
    \label{fig:Comparison_Etot_Hc_MHD_conservation_mean_B0_0_12}
    
\end{figure}

Figure~\ref{fig:Comparison_Etot_Hc_MHD_conservation_mean_B0_0_12} shows the mean temporal evolution of the total energy $E_{\rm tot}$ and the cross helicity $H_c$, averaged over all test sequences. Shaded regions indicate one standard deviation around the mean.\\
The ConvLSTM-UNet model follows the ground-truth trajectories more closely and exhibits a substantially reduced energy drift (table~\ref{tab:invariant_comparison_B0_0_12}), indicating improved long-term conservation of $E_{\rm tot}$.
The cross-helicity error is also consistently lower, suggesting a better preservation of the alignment between the velocity and magnetic fields.
This suggests that the LSTM‑based inductive bias stabilizes the autoregressive rollout and preserves global MHD invariants more effectively on average.

The sign of $H_c$ indicates the dominant alignment between velocity and magnetic field:
$H_c>0$ corresponds to preferential alignment ($\mathbf{u}$ and $\mathbf{B}$ mostly point in the same direction),
whereas $H_c<0$ corresponds to preferential anti-alignment (opposite directions).
Values close to $H_c\approx 0$ indicate either weak global alignment or cancellations between aligned and anti-aligned regions.

\begin{figure}
\includegraphics[width=1.05\linewidth]{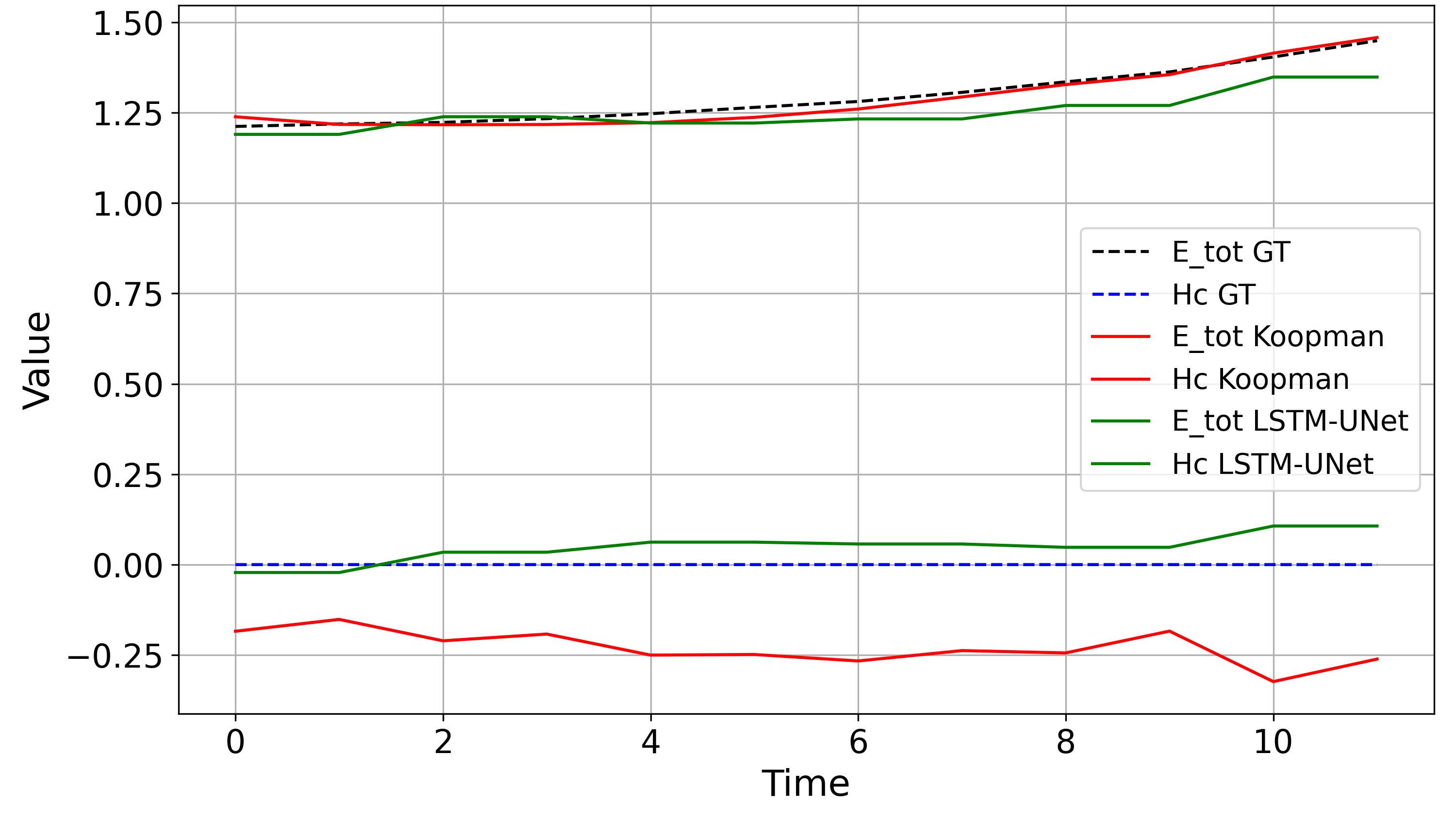}
\caption{\label{fig:KT_LSTM_Unet_comparison_invariants_wj_seq38_B0_0_12} 
Evolution of total energy $E_{tot}$ and cross helicity $H_c$ of both models and the ground truth for a specific sequence for  $B_0=[0.12,0]$.}
\end{figure}

Table~\ref{tab:invariant_comparison_B0_0_12} compares the global conservation metrics of MHD invariants.
The ConvLSTM-UNet model achieves a mean relative error in the total energy $E_{\rm tot}$ of $1.24\%$, which is approximately $6.5\times$ lower than that of the Koopman Transformer ($8.10\%$).
The energy drift is similarly reduced by a factor of $\sim 3.3$, while the root mean squared error of the cross helicity $H_c$ is $2.9\times$ smaller.
Overall, these results indicate that ConvLSTM-UNet preserves MHD invariants more accurately on average.\\
On the other hand, sequence‑level plots shown in figure~\ref{fig:KT_LSTM_Unet_comparison_invariants_wj_seq38_B0_0_12} reveal that Koopman occasionally matches the ground truth more closely in particular sequences, while LSTM‑UNet remains superior on average. These exceptions indicate a regime‑dependent behavior: Koopman may capture certain coherent structures better, whereas LSTM‑UNet provides more reliable conservation across the full ensemble. This motivates reporting both mean statistics and representative sequences.\\
The mean results establish LSTM‑UNet as the more physically stable model, but the per‑sequence analysis highlights that Koopman can outperform in specific cases. This suggests that Koopman’s latent linear dynamics can be advantageous for certain flow regimes, while LSTM‑UNet better controls cumulative error over long horizons. The overall conclusion is therefore not one model dominates everywhere, but rather LSTM‑UNet is more robust globally, with Koopman excelling in a subset of sequences.

\begin{table}[h]
\centering
\caption{Global invariant conservation metrics for Koopman-Transformer and ConvLSTM-UNet models on the Kelvin–-Helmholtz test set ($B_0=0.12$, horizon=2, rollout=6). All errors are averaged over all sequences.}
\label{tab:invariant_comparison_B0_0_12}
\begin{tabular}{lcc}
\hline\hline
\textbf{Metric} & \textbf{Koopman Transformer} & \textbf{ConvLSTM-UNet} \\
\hline
Mean relative error $E_{\mathrm{tot}}$ (\%) & 8.10 & 1.24 \\
Mean energy drift (\%)                      & 17.55 & 5.25 \\
RMSE $H_c$ ($\times10^{-3}$)               & 4.34 & 1.50 \\
\hline\hline
\end{tabular}
\end{table}

\begin{figure*}[ht]
    \centering

    \begin{subfigure}{0.49\linewidth}
        \centering
        \includegraphics[width=\linewidth]{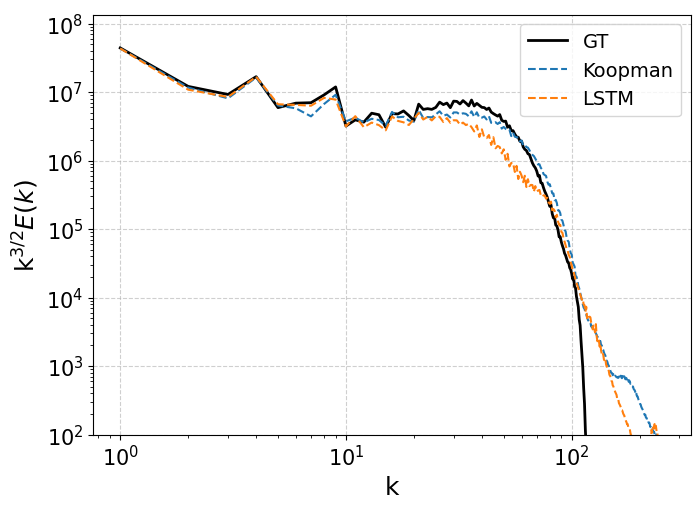}
    \end{subfigure}
    \hfill
    \begin{subfigure}{0.49\linewidth}
        \centering
        \includegraphics[width=\linewidth]{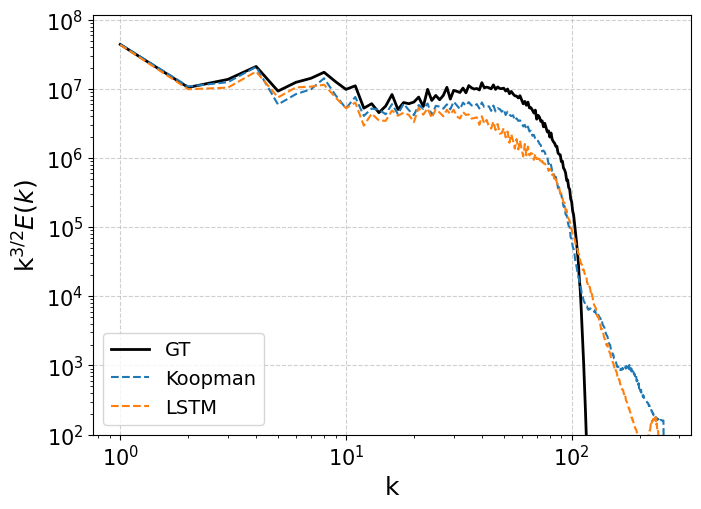}

    \end{subfigure}

    \caption{Compensated total-energy spectra $k^{3/2}E_{\mathrm{tot}}(k)$ at early (t=23, left) and late (t=26, right) rollout steps corresponding to predictions in figures 5 and 6 for Ground Truth (GT), Koopman Transformer, and ConvLSTM-UNet.}

    \label{fig:biskamp_etot_seq40_rollout6_B0_0_12}
\end{figure*}

\subsubsection{Energy spectrum analysis}
\label{subsec:energy_spectrum_analysis}
Figure~\ref{fig:biskamp_etot_seq40_rollout6_B0_0_12} shows the compensated total energy spectra $k^{3/2}E(k)$ obtained from the ground truth (GT), the Koopman transformer model, and the ConvLSTM-UNet model, for both early and late autoregressive rollouts. A flatter profile in the compensated representation indicates closer agreement with the $k^{-3/2}$ scaling expected in the MHD inertial range.

At large scales (low wavenumbers, $k \lesssim 10$), all models closely match the ground truth. This indicates that both approaches accurately capture the dominant large-scale dynamics associated with the Kelvin-Helmholtz instability, including the formation and evolution of coherent vortical structures.

In the intermediate range ($10 \lesssim k \lesssim 70$), the compensated spectra exhibit an approximately flat region, suggesting a scaling behavior close to $E(k) \sim k^{-3/2}$. This is consistent with the Iroshnikov-Kraichnan scaling~\cite{kraichnan1967inertial}. 

Both models reproduce this intermediate spectral behavior reasonably well at early rollout times. The Koopman model, in particular, better preserves the spectral shape across this range, suggesting a more faithful representation of nonlinear energy transfer mechanisms.

At higher wavenumbers, deviations between the models and the ground truth become apparent. While the ground truth spectrum exhibits a sharp decay, corresponding to a numerical cutoff associated with the finite spatial resolution of the simulation rather than as a physical inertial-range behavior of ideal MHD, and is further reinforced by the filtering applied in the CMM code\cite{yin2024characteristic}, both models display a smoother decay and retain excess energy at small scales. This behavior indicates a lack of proper dissipation and the accumulation of spurious high-wavenumber energy.

Multi-step predictions exhibit clear signatures of error accumulation over increasing time horizons. During short rollouts (i.e., the first few timesteps), spectral agreement remains excellent across all scales. As the prediction horizon extends, discrepancies first emerge at high wavenumbers ($k > 70$), where the models retain excess energy compared to the ground truth. With further rollout, this degradation progressively propagates toward intermediate scales.

Overall, these results demonstrate that while both models are capable of capturing large-scale structures and short-term dynamics, their ability to reproduce long-term spectral properties and small-scale physics remains limited. The Koopman-based approach shows improved spectral fidelity compared to the ConvLSTM-UNet model, especially in the inertial range, but both approaches exhibit deficiencies in representing dissipative processes.

These spectral results complement the image-based metrics (SSIM, PSNR) and confirm that both surrogates capture the physically relevant multiscale structure of Kelvin-Helmholtz dynamics in ideal 2D incompressible MHD. While both models accurately reproduce the resolved inertial-range behavior, their differences are primarily confined to the high-$k$ end of the magnetic spectrum, where current-sheet-related structures are most sensitive to numerical and modeling errors.

\subsubsection{Computational efficiency and inference speed}
\label{subsec:InferenceSpeed}

This study demonstrates the potential of autoregressive deep-learning frameworks as fast and physically meaningful surrogates for ideal incompressible MHD.
High-resolution direct numerical simulations of the Kelvin-Helmholtz instability remain computationally expensive: for instance, a simulation at $512 \times 512$ resolution with a snapshot sampling interval $\Delta t_s = 0.5$ (i.e., snapshot time
where the data are stored on the disk) up to $t_{end} = 40$ typically requires on the order of eight hours of wall-clock time.
In contrast, once trained, the neural-network models considered here generate full spatio-temporal trajectories in a matter of minutes, making them well suited for parameter exploration and fast prototyping.

In addition, the computational cost of inference is extremely low. For instance, in the extrapolation experiment with $B_0 = 0.12$, 
44 sequences with 12 predicted frames each (528 frames in total) require only 18.75 seconds of computation, 
corresponding to roughly $3.5\times10^{-2}$ seconds per frame. 
Similarly, for the interpolation case with $B_0 = 0.08$, 
64 sequences totaling 768 frames are predicted in 25.41 seconds, 
or about $3.3\times10^{-2}$ seconds per frame. 
Each prediction simultaneously reconstructs both the vorticity and current density fields. 
For comparison, a direct numerical simulation of the Kelvin-Helmholtz instability at $B_0 = 0.05$ requires approximately 8 hours to generate 101 snapshots, corresponding to an average computational cost of about 285 seconds per snapshot. 

Comparing these values shows that the trained machine learning models produce predictions roughly four orders of magnitude faster than DNS, corresponding to a speed-up of approximately $8\times10^3$ at inference time.

\begin{table*}[ht]
\centering
\caption{Comparison of KT-MHD Transformer and ConvLSTM-UNet based on mean MSE, RMSE, MAE, SSIM, and PSNR metrics, computed by averaging predictions across all sequences and the entire autoregressive rollouts for $B_0 = [0.12,0]$.}
\label{tab:metrics_comparison_B0_0_12}
\resizebox{\textwidth}{!}{%
\begin{tabular}{lcccccccc}
\hline
Model & Quantities & MSE & RMSE & MAE & SSIM & PSNR (dB) & Epochs\\ 
\hline
KT-MHD & ($\omega, j$) & (0.011, 0.43) & (0.10, 0.66) & (0.04, 0.10) & (0.77, 0.42) & (33.62, 23.67) & 500 \\
ConvLSTM-UNet & ($\omega, j$) & (0.009, 0.59) & (0.09, 0.77) & (0.036, 0.10) & (0.78, 0.47) & (35.37, 24.56) & 500 \\ 
\hline
\end{tabular}}

\end{table*}

\section{Conclusion and Discussion}
\label{sec:Conclusion}

In this work, we investigated data-driven autoregressive forecasting of the two-dimensional incompressible ideal MHD applied to the Kelvin-Helmholtz instability for different magnetic intensities. We compared two representative neural architectures: a convolutional recurrent ConvLSTM-UNet and a Koopman-based Transformer (KT-MHD), which respectively emphasize local spatio-temporal modeling and global attention-driven latent linear dynamics.
Both models successfully reproduce the nonlinear development of the instability and generate stable multi-step autoregressive predictions, capturing vortex roll-up, current-sheet formation, and large-scale flow evolution.

A detailed comparison reveals complementary strengths and limitations.
The ConvLSTM-UNet benefits from local convolutional refinement and multi-scale skip connections, yielding sharper reconstructions and superior image-based scores such as PSNR and SSIM, particularly for the vorticity field.
In contrast, the KT-MHD Transformer exhibits improved spectral consistency, and demonstrating enhanced long-range temporal coherence.

The spectral comparison, based on the total rescaled energy spectrum $k^{3/2} E_{\mathrm{tot}}(k)$, demonstrates that both neural network models successfully capture the main features of magnetohydrodynamic energy cascade dynamics at large and intermediate scales, confirming the learnability of turbulent energy transfer mechanisms. 
Large and intermediate scales are generally well predicted, reflecting the dominant coherent structures and the ability of the models to approximate energy transfer across the cascade. In contrast, small scales constitute the primary source of prediction error, due both to the inherently chaotic nature of turbulence.

Beyond spectral diagnostics, the conservation of global MHD invariants provides a physics-based assessment of long-term forecast reliability. 
Tracking the total energy $E_{\rm tot}$ and cross helicity $H_c$ during autoregressive rollouts provides additional insight into the ability of the models to preserve key physical properties.
The ConvLSTM-UNet exhibits lower energy drift and reduced cross-helicity error on average, indicating improved preservation of ideal MHD constraints over long horizons. In contrast, the Koopman Transformer shows larger mean deviations and drift, suggesting a more regime-dependent behavior. These results highlight the importance of invariant-based diagnostics, which capture aspects of physical fidelity not reflected by image-based metrics alone.

Part of the KT-MHD Transformer’s performance can be attributed to its use of variable-specific decoders, which allow a more structured representation of the coupled MHD variables. This design choice reflects a trade-off between physical specialization and computational cost rather than an attempt to favor one architecture, as introducing multiple decoders in the ConvLSTM-UNet would substantially increase its already high training cost.

These differences also translate into markedly different computational costs: as reported in Table~\ref{tab:training_cost}, the KT-MHD Transformer requires approximately $1.5$ hours of GPU training time, compared to about $7$ hours for the ConvLSTM-UNet under identical training conditions, reflecting a nearly fivefold reduction in training cost.

Overall, neither architecture dominates across all diagnostics.
The ConvLSTM-UNet provides greater robustness in terms of global invariant conservation and average long-term stability, while the Koopman Transformer excels in preserving multiscale spectral content and capturing coherent magnetic structures in specific regimes.
These complementary strengths suggest that hybrid approaches combining global attention-based modeling with local convolutional refinement may offer a promising direction for future MHD surrogate models.

Notably, these results are obtained despite a relatively limited training dataset, restricted to a finite set of magnetic field strengths and initial conditions.
The ability of both architectures to reproduce multiscale structures, preserve key physical invariants, and maintain stable long-term autoregressive rollouts indicates a strong degree of robustness and generalization.
This suggests that the learned representations capture essential dynamical features of the underlying ideal MHD system rather than merely interpolating within the training data.

As discussed in subsection~\ref{subsec:InferenceSpeed}, this substantial acceleration highlights the potential of deep-learning-based surrogate models for rapid exploration of MHD parameter spaces. Once trained on a limited set of simulations, the models can efficiently generate new temporal evolutions for different magnetic field strengths without the need to perform additional expensive DNS computations. While DNS remains essential for producing physically consistent reference data, surrogate models provide a powerful complementary tool for fast prediction, parameter exploration, and large-scale data generation.

Future work may focus on incorporating physics-informed constraints directly into the training objective, such as enforcing divergence-free fields or penalizing violations of MHD invariants.
Extending the framework to more complex plasma models, such as the Hasegawa--Wakatani system~\cite{lin2024synthesizing}, represents another promising direction to further assess the generality and robustness of autoregressive neural network surrogates.

\begin{acknowledgments}

The authors gratefully acknowledge the support of ECE (Engineering School) in Paris, which provides funding for the doctoral research of David Kivarkis. This financial support has made this work possible and is sincerely appreciated.
The authors were granted access to the HPC resources of  Centre de Calcul Intensif d’Aix-Marseille Universit{\'e}.
We acknowledge the financial support from the French Federation for Magnetic Fusion Studies (FR-FCM) and the Eurofusion consortium, funded by the Euratom Research and Training Programme under Grant Agreement No. 633053. The views and opinions expressed herein do not necessarily reflect those of the European Commission.

\end{acknowledgments}

\section*{Data Availability Statement}
The data that support the findings of this study are available from the corresponding author upon reasonable request.

\appendix*
\section{Assessment of autoregressive model predictions in the interpolation regime
}
\setcounter{figure}{0}
\setcounter{table}{0}
\renewcommand{\thefigure}{A.\arabic{figure}}
\renewcommand{\thetable}{A.\arabic{table}}

While the main body of the paper focuses on the autoregressive forecasting of vorticity $\omega$ and current density $j$ for an initial magnetic field $B_0 = [0.12, 0]$, this appendix reports additional results for $B_0 = [0.08, 0]$, corresponding to an unseen magnetic field strength that lies between the values used during training. 
These additional experiments are included to assess the robustness and generalization capability of the Koopman Transformer and ConvLSTM-UNet models.\\
In particular, we provide both qualitative and quantitative results of the Kelvin-Helmholtz instability predictions for vorticity and current density fields. 
Visual prediction snapshots are shown at two representative stages: close to the onset of the instability and at a later nonlinear phase. 
This approach allows us to assess how accurately each model captures the early development of shear-driven structures as well as their subsequent nonlinear evolution.

As expected, both models exhibit improved stability and accuracy compared to the
extrapolative setting. The large-scale Kelvin-Helmholtz billow, its roll-up, and the global phase of the instability are well reproduced throughout the autoregressive
rollout for both vorticity $\omega$ and current density $j$.

\begin{figure*}
\includegraphics[width=1\linewidth]{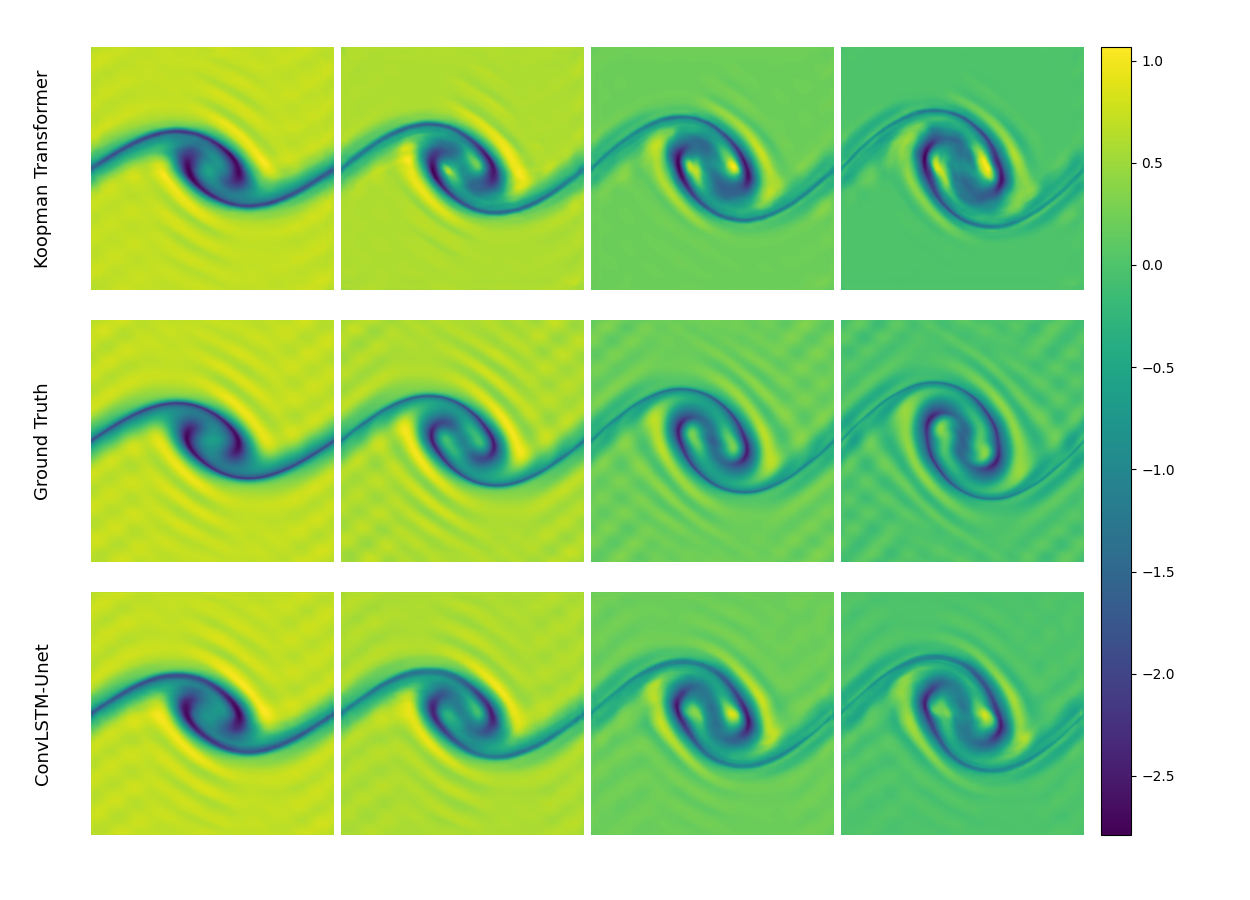}
\caption{\label{fig:vorticity_prediction_KT_LSTM_Unet_B0_0_08_seq40} Autoregressive forecasting of the vorticity field $\omega$ for $B_0=[0.08,0]$. 
    From top to bottom: KT-MHD Transformer prediction, Ground Truth (DNS reference), and ConvLSTM U-Net prediction. For visualization purposes, only every third frame is shown, corresponding to time steps $t = 23, 23.5, 24, \dots, 28.5$.}
\end{figure*}

\begin{figure*}
\includegraphics[width=1\linewidth]{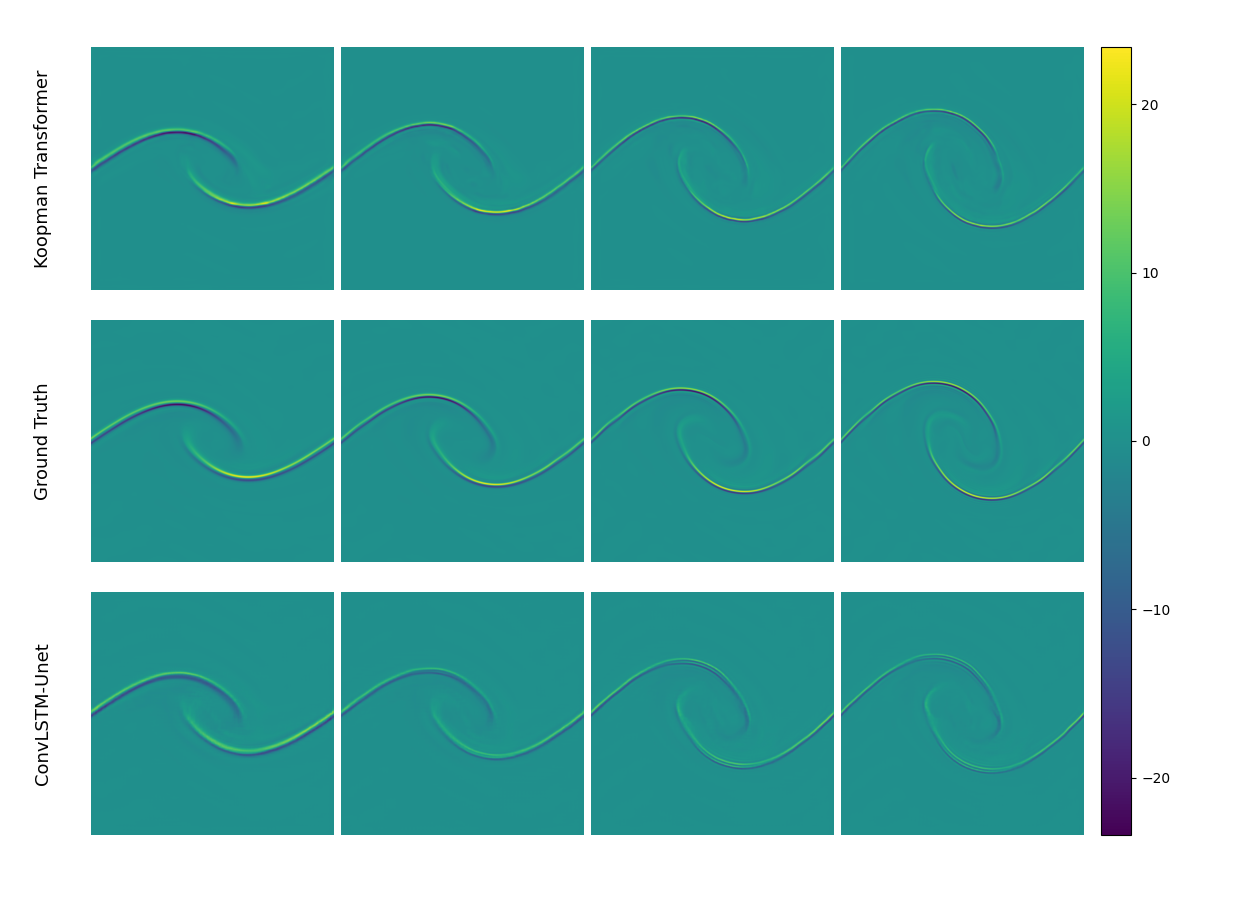}
\caption{\label{fig:current_prediction_KT_LSTM_Unet_B0_0_08_seq40} Autoregressive forecasting of the current density field $j$ for $B_0=[0.08,0]$. 
    From top to bottom: KT-MHD Transformer prediction, Ground Truth (DNS reference), and ConvLSTM U-Net prediction. For visualization purposes, only every third frame is shown, corresponding to time steps $t = 23, 23.5, 24, \dots, 28.5$.}
\end{figure*}

\begin{figure*}
\includegraphics[width=1\linewidth]{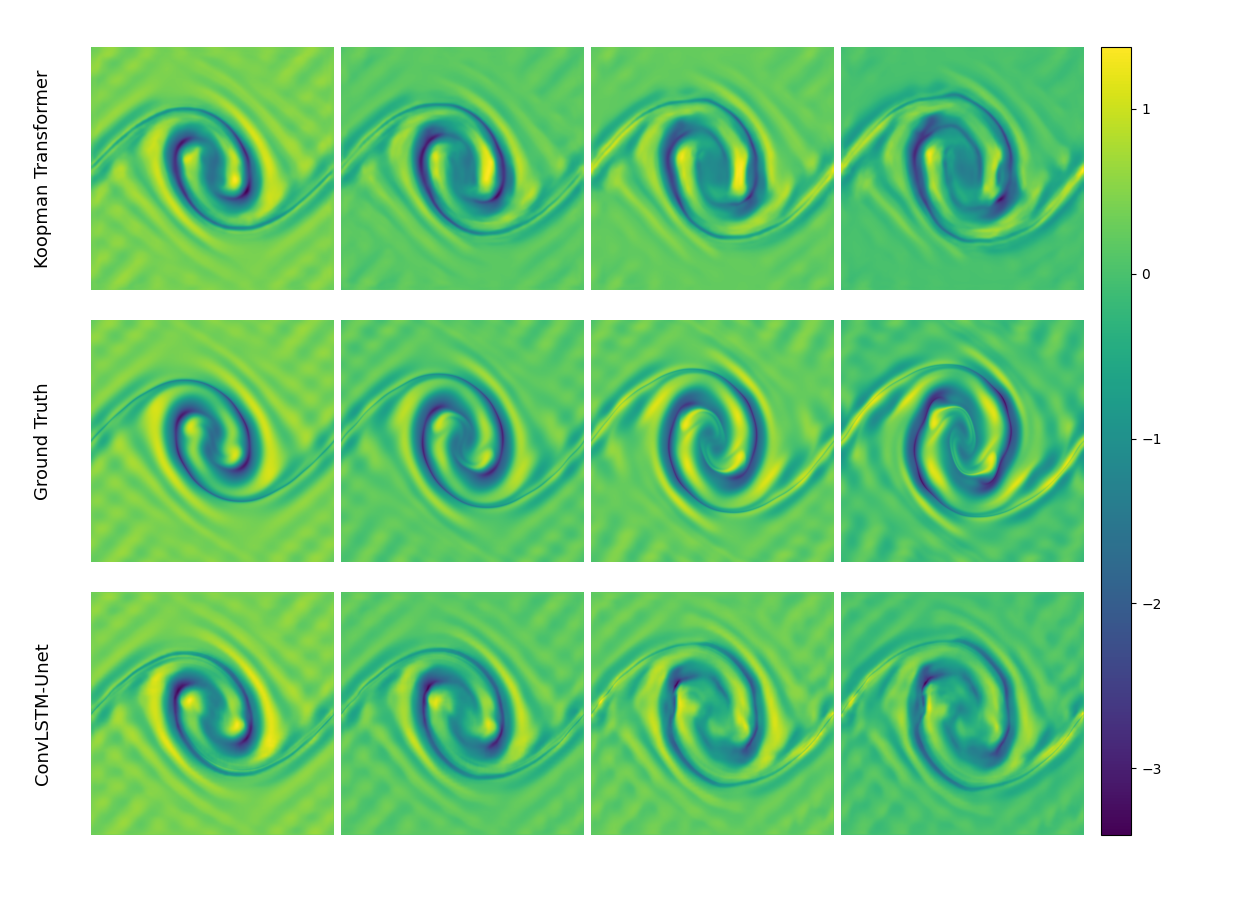}
\caption{\label{fig:vorticity_prediction_KT_LSTM_Unet_B0_0_08_seq50} Autoregressive forecasting of the vorticity field $\omega$ for $B_0=[0.08,0]$, at a later stage of the Kelvin–-Helmholtz instability.
    From top to bottom: KT-MHD Transformer prediction, Ground Truth (DNS reference), and ConvLSTM U-Net prediction. For visualization purposes, only every third frame is shown, corresponding to time steps $t = 28, 28.5, 29, \dots, 33.5$.}
\end{figure*}

\begin{figure*}
\includegraphics[width=1\linewidth]{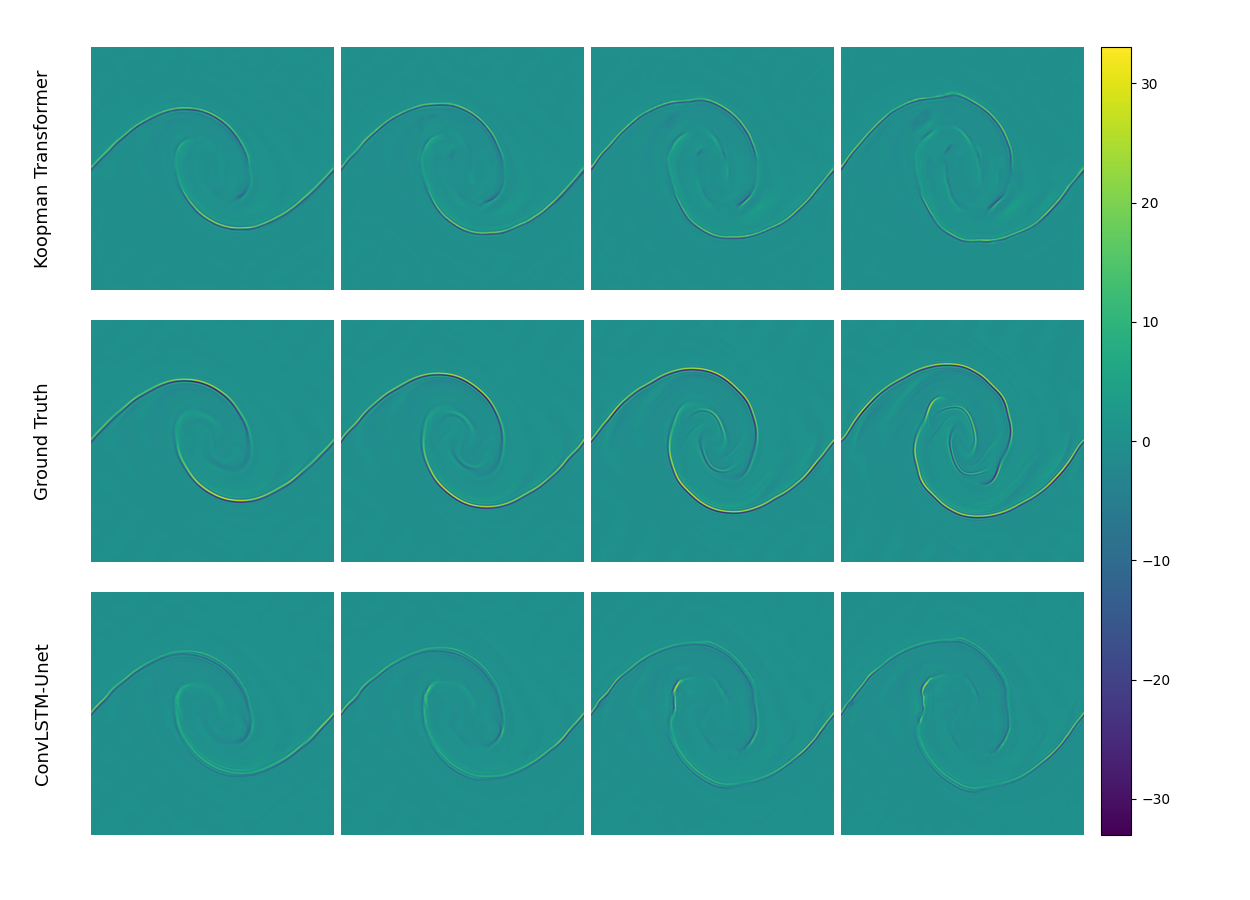}
\caption{\label{fig:current_prediction_KT_LSTM_Unet_B0_0_08_seq50} Autoregressive forecasting of the current density field $j$ for $B_0=[0.08,0]$, at a later stage of the Kelvin–-Helmholtz instability. 
    From top to bottom: KT-MHD Transformer prediction, Ground Truth (DNS reference), and ConvLSTM U-Net prediction. For visualization purposes, only every third frame is shown, corresponding to time steps $t = 28, 28.5, 29, \dots, 33.5$.}
\end{figure*}

The visual patterns observed for $B_0 = [0.08, 0]$ in figures~\ref{fig:vorticity_prediction_KT_LSTM_Unet_B0_0_08_seq40}, \ref{fig:current_prediction_KT_LSTM_Unet_B0_0_08_seq40}, \ref{fig:vorticity_prediction_KT_LSTM_Unet_B0_0_08_seq50}, and \ref{fig:current_prediction_KT_LSTM_Unet_B0_0_08_seq50} are similar to those described in the main text for $B_0 = [0.12, 0]$, indicating that the models reproduce comparable qualitative behavior across magnetic field strengths.
The ConvLSTM-UNet preserves sharper vorticity gradients
and finer-scale rotational structures, while the Koopman-Transformer yields smoother
velocity fields but demonstrates increased robustness in the prediction of the current density, maintaining coherent current sheets and realistic magnetic structures over time. Compared to extrapolation, these differences are less pronounced, reflecting the reduced difficulty of the interpolative regime.

\begin{figure*}[ht]
    \centering

    \begin{subfigure}{0.49\linewidth}
        \centering
        \includegraphics[width=\linewidth]{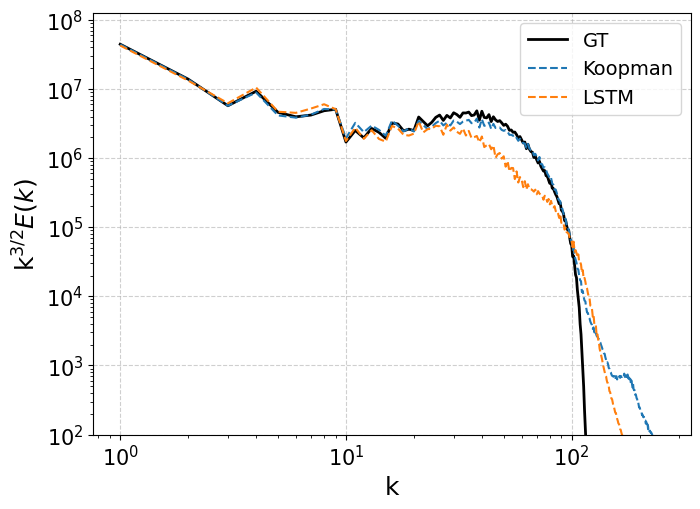}
    \end{subfigure}
    \hfill
    \begin{subfigure}{0.49\linewidth}
        \centering
        \includegraphics[width=\linewidth]{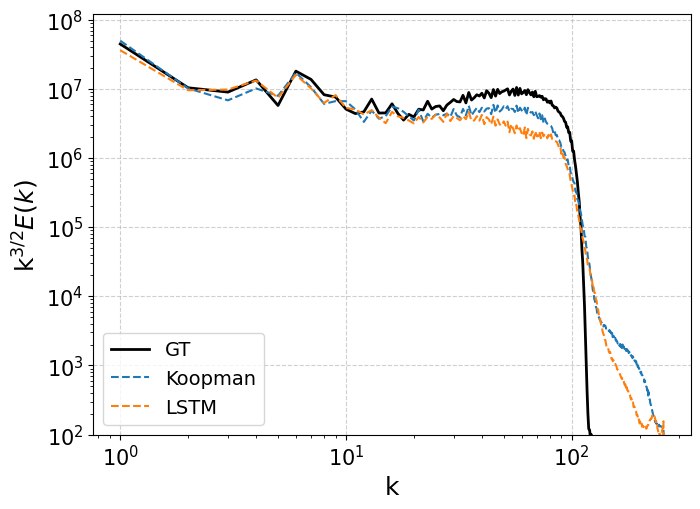}

    \end{subfigure}

    \caption{Comparison of compensated total-energy spectra $k^{3/2}E_{\mathrm{tot}}(k)$ at early (t=25, left) and late (t=28, right) rollout steps corresponding to predictions in figures A.2 and A.3, for Ground Truth, Koopman Transformer, and ConvLSTM-UNet.}
    \label{fig:biskamp_etot_seq40_rollout6_B0_0_08}
\end{figure*}

\begin{figure*}[ht]
    \centering

    \begin{subfigure}{0.49\linewidth}
        \centering
        \includegraphics[width=\linewidth]{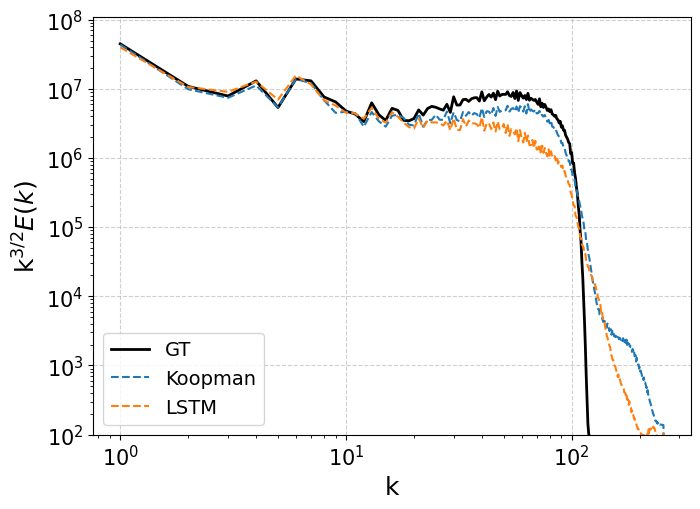}

    \end{subfigure}
    \hfill
    \begin{subfigure}{0.49\linewidth}
        \centering
        \includegraphics[width=\linewidth]{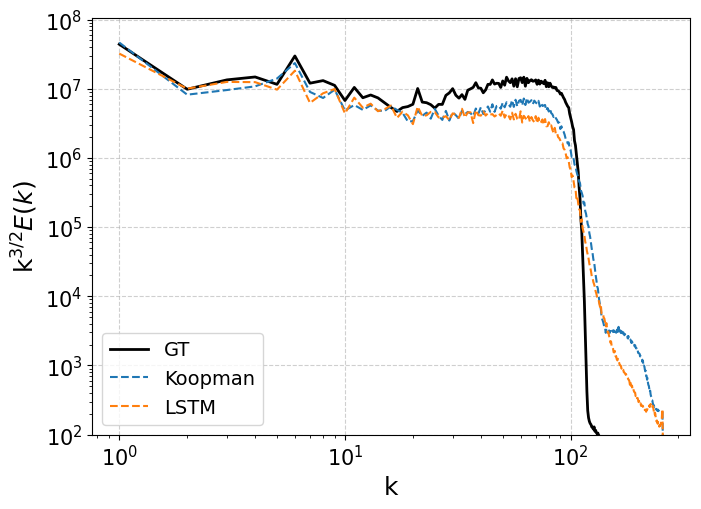}

    \end{subfigure}

    \caption{Comparison of compensated total-energy spectra $k^{3/2}E_{\mathrm{tot}}(k)$ at early (t=28.5, left) and late (t=30.5, right) rollout steps corresponding to predictions in figures A.5 and A.6, for Ground Truth, Koopman Transformer, and ConvLSTM-UNet, at a later stage of the Kelvin-Helmholtz instability.}

    \label{fig:biskamp_etot_seq50_rollout6_B0_0_08}
\end{figure*}

The compensated spectra shown in figures~\ref{fig:biskamp_etot_seq40_rollout6_B0_0_08}, and \ref{fig:biskamp_etot_seq50_rollout6_B0_0_08} further support these observations. 
Both models reproduce the large and intermediate scale kinetic and magnetic spectra with good fidelity, closely matching the DNS reference.

These qualitative and spectral observations are further supported by the quantitative metrics summarized in Tables~\ref{tab:koopman_vorticity_metrics_B0_0_08_seq40},~\ref{tab:lstmunet_vorticity_metrics_B0_0_08_seq40},~\ref{tab:koopman_current_density_metrics_B0_0_08_seq40}, and ~\ref{tab:lstmunet_current_density_metrics_B0_0_08_seq40}. The ConvLSTM-UNet achieves higher SSIM and PSNR for the vorticity field, while the Koopman-Transformer exhibits lower RMSE and competitive MAE for the current density. As expected, the overall metric values are slightly better than in the extrapolative case, reflecting the reduced difficulty of predicting within the training parameter range. And the global performance metrics listed in Table~\ref{tab:metrics_comparison_B0_08} reinforce these observations, confirming that both models achieve similar accuracy in reconstructing vorticity and current density.

\begin{table}[ht]
\centering
\caption{Performance metrics for the Koopman Transformer model on the predicted frames for the vorticity field shown in figure~\ref{fig:vorticity_prediction_KT_LSTM_Unet_B0_0_08_seq40}.}
\label{tab:koopman_vorticity_metrics_B0_0_08_seq40}
\renewcommand{\arraystretch}{1.7}
\resizebox{0.9\linewidth}{!}{
\begin{tabular}{c c c c c c}
\hline
Timestep t & MSE & RMSE & MAE & SSIM & PSNR \\
\hline
23 & 0.0005 & 0.023 & 0.015 & 0.914 & 38.63 \\
24.5 & 0.003 & 0.055 & 0.036 & 0.726 & 33.05 \\
26 & 0.010 & 0.101 & 0.058 & 0.673 & 30.08 \\
27.5 & 0.026 & 0.162  & 0.092  & 0.597  & 27.51  \\
\hline
\end{tabular}}
\end{table}

% Tableau suivant pour ConvLSTM-UNet
\begin{table}[ht]
\centering
\caption{Performance metrics for the ConvLSTM-UNet model on the predicted frames for the vorticity field shown in figure~\ref{fig:vorticity_prediction_KT_LSTM_Unet_B0_0_08_seq40}}
\label{tab:lstmunet_vorticity_metrics_B0_0_08_seq40}
\renewcommand{\arraystretch}{1.7}
\resizebox{0.9\linewidth}{!}{
\begin{tabular}{c c c c c c}
\hline
Timestep t & MSE & RMSE & MAE & SSIM & PSNR \\
\hline
23 & 0.0016 & 0.041 & 0.020 & 0.941 & 33.84 \\
24.5 & 0.0047 & 0.069 & 0.039 & 0.797 & 30.24 \\
26 & 0.0070 & 0.084 & 0.052 & 0.722 & 30.84 \\
27.5 & 0.015   & 0.126  & 0.082  & 0.613  & 28.16  \\
\hline
\end{tabular}}
\end{table}

\begin{table}[ht]
\centering
\caption{Performance metrics for the Koopman Transformer model on the predicted frames for the current density field shown in figure~\ref{fig:current_prediction_KT_LSTM_Unet_B0_0_08_seq40}. }
\label{tab:koopman_current_density_metrics_B0_0_08_seq40}
\renewcommand{\arraystretch}{1.7}
\resizebox{0.9\linewidth}{!}{
\begin{tabular}{c c c c c c}
\hline
Timestep t & MSE & RMSE & MAE & SSIM & PSNR \\
\hline
23 & 0.022 & 0.149 & 0.057 & 0.919 & 38.97 \\
24.5 & 0.076 & 0.277 & 0.093 & 0.925 & 37.93 \\
26 & 0.267 & 0.517 & 0.161 & 0.914 & 35.95 \\
27.5 & 1.432   & 1.197  & 0.311  & 0.875  & 30.09  \\
\hline
\end{tabular}}
\end{table}

% Tableau suivant pour ConvLSTM-UNet
\begin{table}[ht]
\centering
\caption{Performance metrics for the ConvLSTM-UNet model on the predicted frames for the current density field shown in figure~\ref{fig:current_prediction_KT_LSTM_Unet_B0_0_08_seq40}} 
\label{tab:lstmunet_current_density_metrics_B0_0_08_seq40}
\renewcommand{\arraystretch}{1.7}
\resizebox{0.9\linewidth}{!}{
\begin{tabular}{c c c c c c}
\hline
Timestep t & MSE & RMSE & MAE & SSIM & PSNR \\
\hline
23 & 0.136 & 0.369 & 0.085 & 0.919 & 28.44 \\
24.5 & 0.442 & 0.665 & 0.151 & 0.895 & 25.40 \\
26 & 1.140 & 1.068 & 0.233 & 0.901 & 26.89 \\
27.5 & 2.592   & 1.610  & 0.373  & 0.877  & 26.35  \\
\hline
\end{tabular}}
\end{table}

While the preceding results demonstrate accurate reconstruction of flow structures and spectra, the preservation of global MHD invariants provides a complementary measure of long-term physical stability.

\begin{figure*}[ht]
    \centering

    % --- Colonne 1 ---
    \begin{subfigure}{0.48\linewidth}
        \centering
        \includegraphics[width=\linewidth]{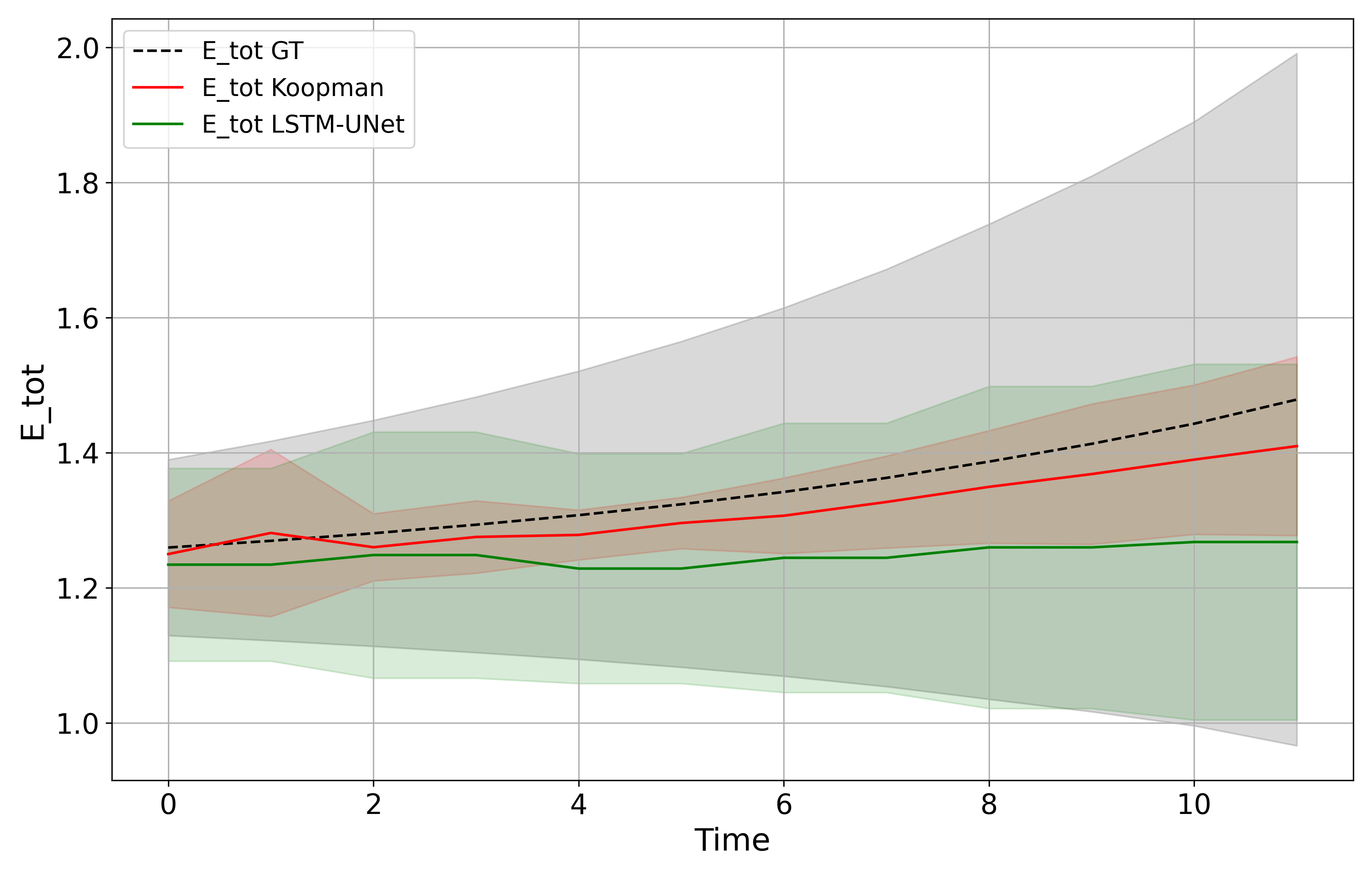}
        \caption{Evolution of the total energy ($E_{tot}$) over time for both models against the ground truth.}
    \end{subfigure}
    \hfill
    % --- Colonne 2 ---
    \begin{subfigure}{0.48\linewidth}
        \centering
        \includegraphics[width=\linewidth]{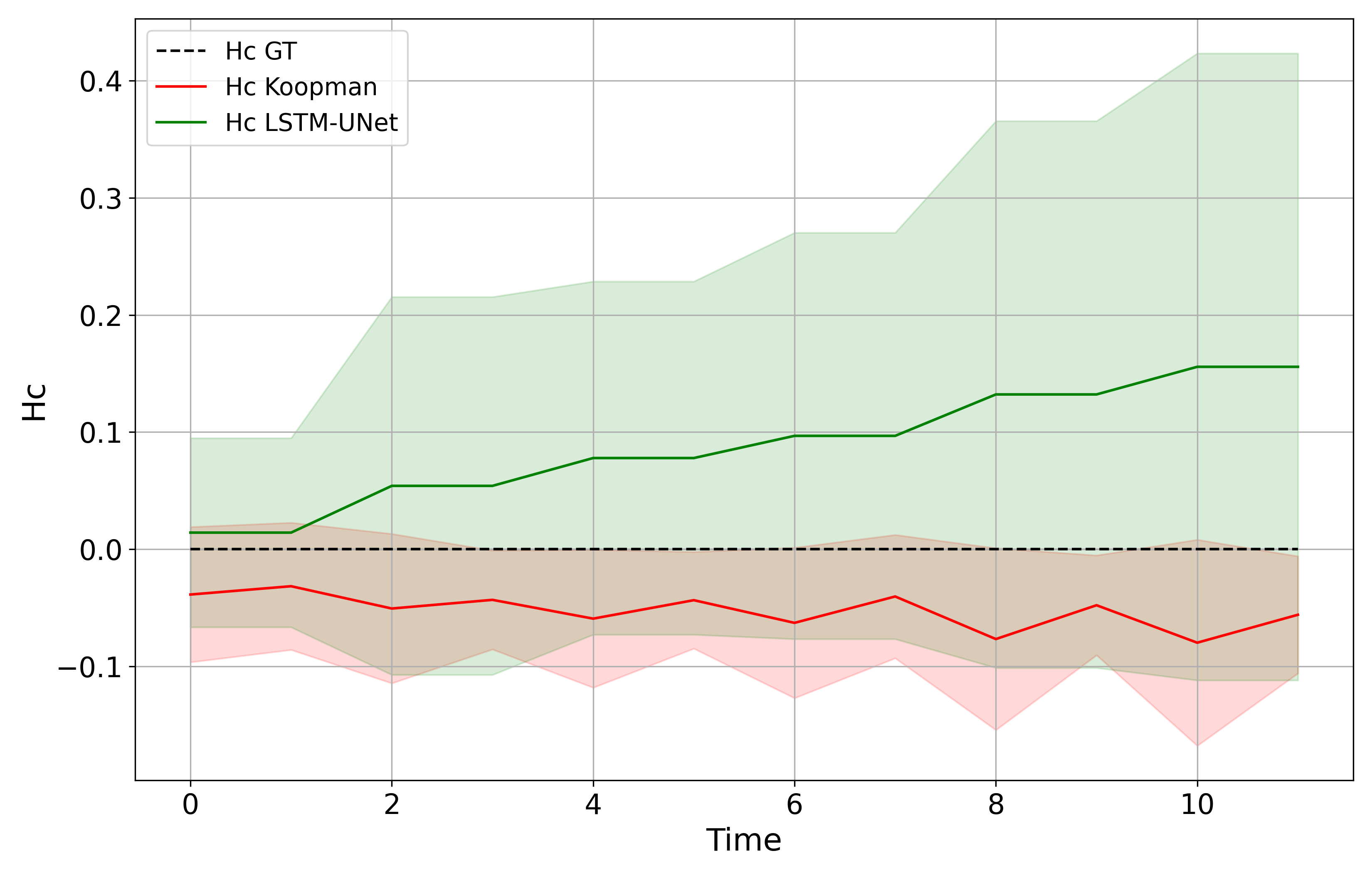}
        \caption{Evolution of the cross helicity ($H_c$) over time for both models against the ground truth.}
    \end{subfigure}

    \caption{Evolution of th emean of MHD invariants, for an initial magnetic field $B_0 =[0.08,0]$. The shaded region corresponds to one standard deviation ($\pm1\sigma$) around the mean value.}
    \label{fig:Comparison_Etot_Hc_MHD_conservation_mean_B0_0_08}
\end{figure*}

The conservation of global MHD invariants for the interpolative case is summarized in Table~\ref{tab:invariant_comparison_B0_0_08} and illustrated in Figure~\ref{fig:Comparison_Etot_Hc_MHD_conservation_mean_B0_0_08}, which shows the mean temporal evolution of the total energy $E_{\mathrm{tot}}$ and cross helicity $H_c$ over all test sequences at each predicted timestep.
Both models remain close to the DNS reference throughout the rollout, indicating stable autoregressive behavior and no significant violation of ideal invariants.

On average, the ConvLSTM--UNet exhibits lower relative error and reduced drift in the total energy, whereas the Koopman--Transformer achieves a smaller RMSE for the cross helicity, indicating a more accurate preservation of velocity–magnetic field alignment in this setting. 
This contrasts with the extrapolative case, where the ConvLSTM--UNet dominated both invariants on average.

\begin{figure}
\includegraphics[width=1\linewidth]{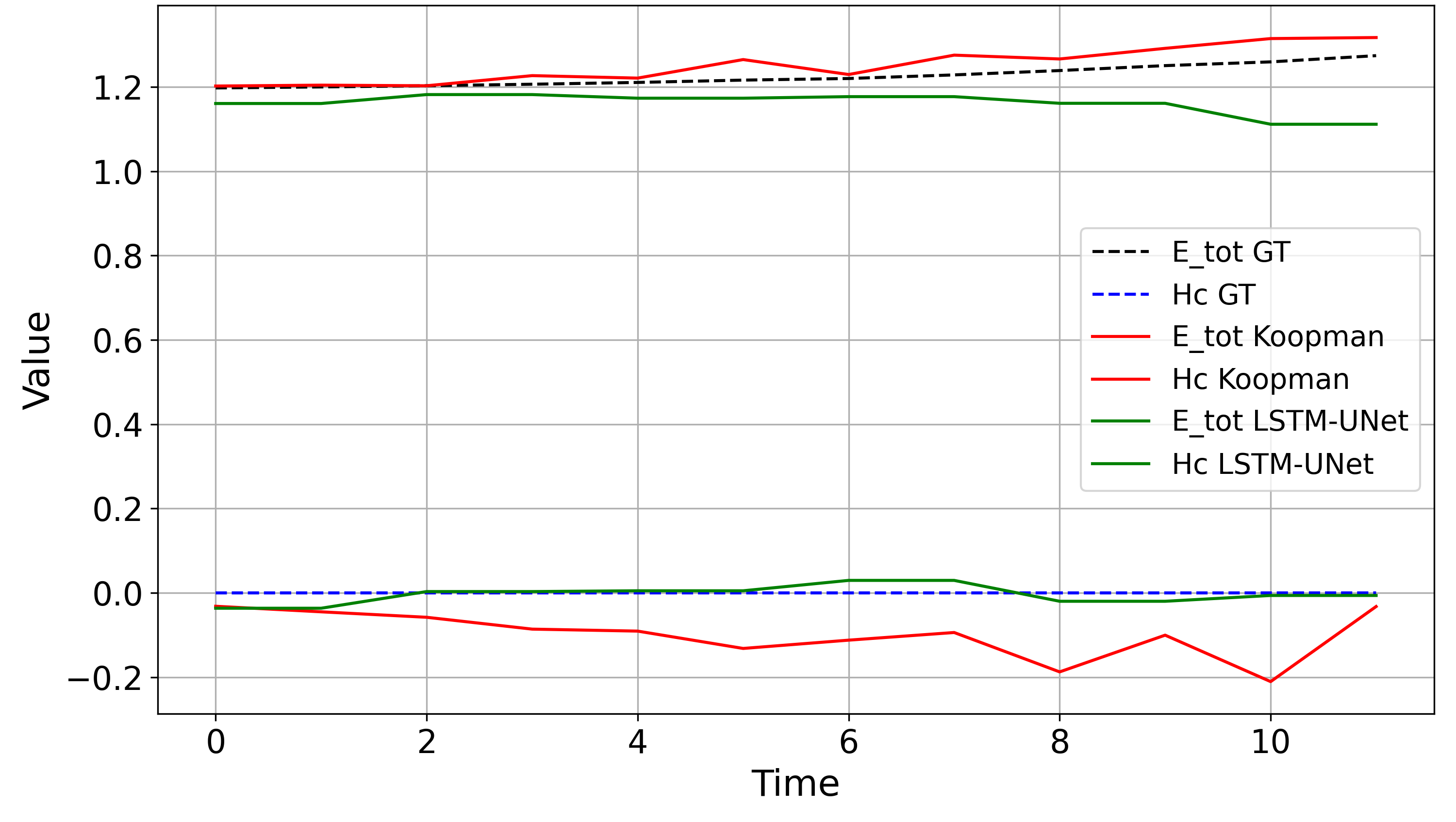}
\caption{\label{fig:KT_LSTM_Unet_comparison_invariants_wj_seq38_B0_0_08} 
Evolution of total energy $E_{tot}$ and cross helicity $H_c$ of both models and the ground truth for a specific sequence for  $B_0=[0.08,0]$.}
\end{figure}

This regime-dependent behavior can be explained by the interpolative nature of the setting, where the dynamics remain within the range of magnetic field strengths seen during training. 
Here, the Koopman--Transformer benefits from its globally coherent latent dynamics, favoring the stable evolution of integrated quantities such as cross helicity, while the ConvLSTM--UNet continues to provide robust energy conservation. 
As in the extrapolative case, sequence-level analysis (Figure~\ref{fig:KT_LSTM_Unet_comparison_invariants_wj_seq38_B0_0_08}) shows that individual sequences can deviate from the average: some sequences favor the ConvLSTM--UNet for cross helicity or energy, illustrating the complementary strengths of the two architectures across different regimes.\\
Overall, the interpolative results confirm that both architectures generalize reliably
within the training parameter range and that the complementary strengths identified in
the extrapolation case persist, while exhibiting reduced error magnitude and improved temporal stability.\\

\begin{table}[ht]
\centering
\caption{Relative improvement of interpolation ($B_0=0.08$) over extrapolation ($B_0=0.12$), averaged over all timesteps. Positive values indicate improved performance in the interpolation regime.}
\label{tab:interpolation_vs_extrapolation}
\renewcommand{\arraystretch}{1.3}
\begin{tabular}{c c c c c}
\hline
Model & Variable & MSE (\%) & SSIM (\%) & PSNR (dB) \\
\hline
Koopman Transformer & $\omega$ & $+50.6$ & $+34.2$ & $+3.8$ \\
Koopman Transformer & $j$      & $+47.8$ & $+8.7$  & $+2.6$ \\
\hline
ConvLSTM-UNet     & $\omega$ & $+38.5$ & $+21.1$ & $+2.4$ \\
ConvLSTM-UNet     & $j$      & $-6.3$  & $+3.9$  & $-0.8$ \\
\hline
\end{tabular}
\end{table}

Table~\ref{tab:interpolation_vs_extrapolation} reports the relative performance variations between the extrapolation regime ($B_0 = 0.12$) to the interpolation regime ($B_0 = 0.08$), using the former as a baseline. For each metric, values are first computed at every autoregressive timestep and then averaged over all timesteps. The relative improvement in MSE is computed as $(M_{\mathrm{extra}} - M_{\mathrm{interp}})/M_{\mathrm{extra}} \times 100$, while SSIM variations are computed analogously. PSNR variations are reported as absolute differences in dB. Here, $M_{\mathrm{extra}}$ and $M_{\mathrm{interp}}$ denote the timestep-averaged metrics in the extrapolation and interpolation regimes, respectively.
A general improvement is observed in the interpolation case across both physical variables and models. For instance, the MSE decreases by up to $\sim 50\%$ for the vorticity field and $\sim 48\%$ for the current density in the Koopman Transformer case, while similar trends are observed for the ConvLSTM-UNet. Improvements are also observed for the ConvLSTM-UNet, although their magnitude varies depending on the variable. These trends are further reflected in the SSIM and PSNR metrics, indicating enhanced pointwise accuracy and structural reconstruction in the interpolation regime.\\

\begin{figure*}[ht]
    \centering

    % -- Ligne 1 --
    \begin{subfigure}{0.9\linewidth}
        \centering
        \includegraphics[width=0.9\linewidth]{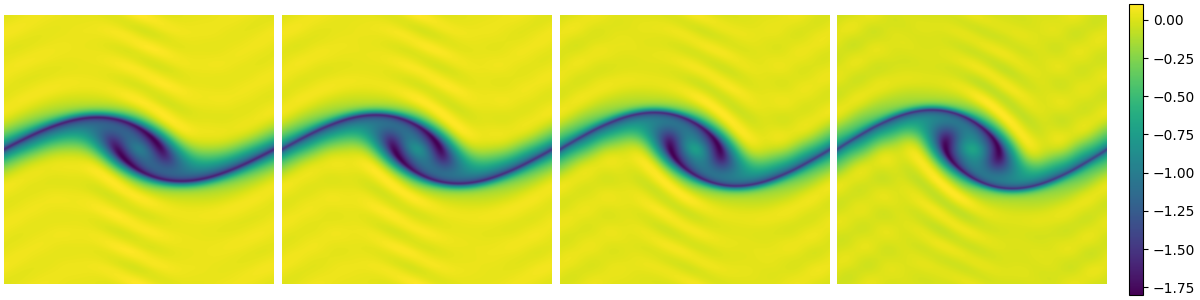}
        \caption{Input sequence for the vorticity field $\omega$}
    \end{subfigure}

    \vspace{1em}

    % -- Ligne 2 --
    \begin{subfigure}{0.9\linewidth}
        \centering
        \includegraphics[width=0.9\linewidth]{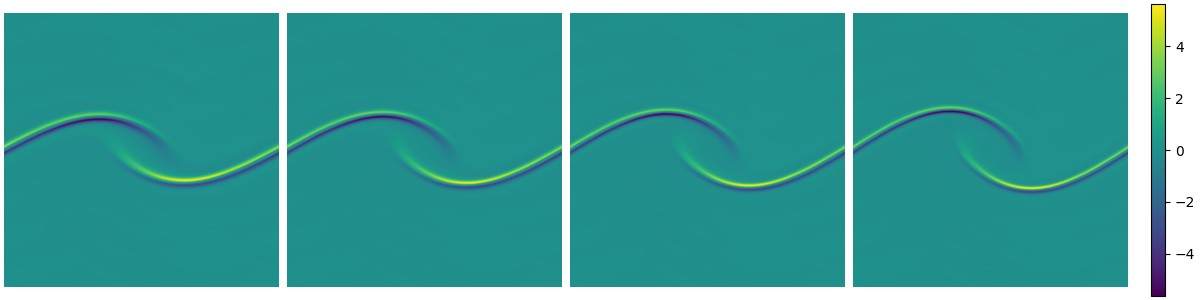}
        \caption{Input sequence for the current density field $j$}
    \end{subfigure}

    \caption{Input sequences of four past frames for  vorticity $\omega$ and current density $j$, with an initial magnetic field $\mathbf{B}_0 = [0.08, 0]$. The inputs correspond to consecutive time steps $t = 21, 21.5, 22, 22.5$.}
    \label{fig:WJ_input_seq40_B0_0_08}
\end{figure*}

\begin{figure*}[ht]
    \centering

    % -- Ligne 1 --
    \begin{subfigure}{0.95\linewidth}
        \centering
        \includegraphics[width=0.9\linewidth]{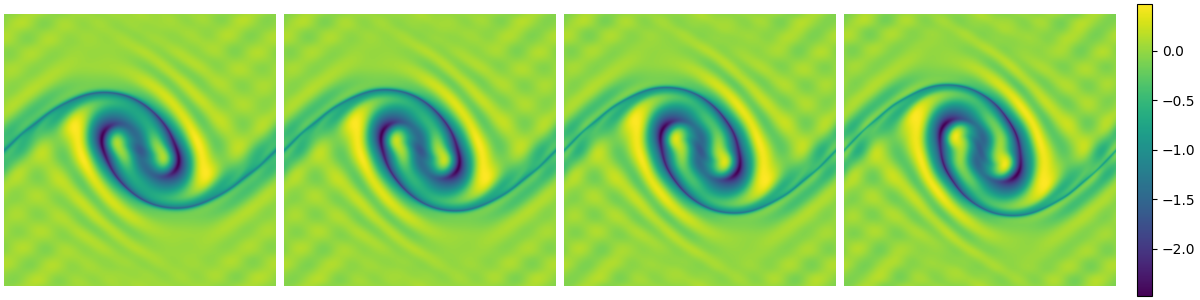}
        \caption{Input sequence for the vorticity field $\omega$}
    \end{subfigure}

    \vspace{1em}

    % -- Ligne 2 --
    \begin{subfigure}{0.95\linewidth}
        \centering
        \includegraphics[width=0.9\linewidth]{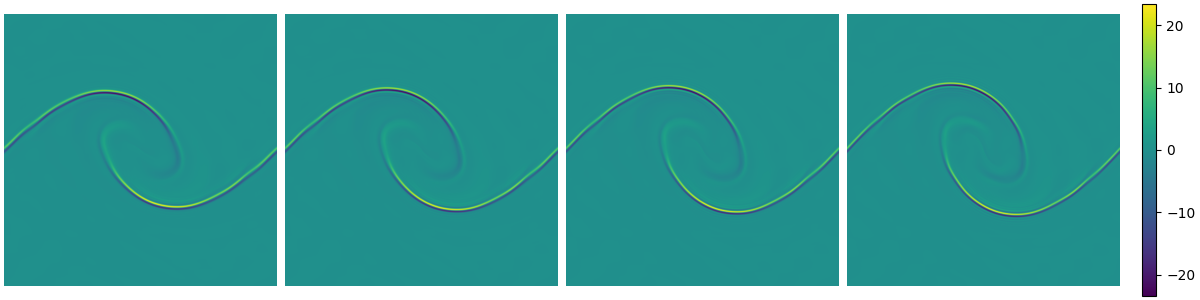}
        \caption{Input sequence for the current density field $j$}
    \end{subfigure}

    \caption{Input sequences of four past frames for  vorticity $\omega$ and current density $j$ , with an initial magnetic field $\mathbf{B}_0 = [0.08, 0]$, at a later stage of the Kelvin-Helmholtz instability. The inputs correspond to consecutive time steps $t = 26, 26.5, 27, 27.5$.}
    \label{fig:WJ_input_seq50_B0_0_08}
\end{figure*}

\begin{table*}[ht]
\centering
\caption{Comparison of KT-MHD Transformer and ConvLSTM-UNet based on mean MSE, RMSE, MAE, SSIM, and PSNR metrics, computed by averaging predictions across all sequences and the entire autoregressive rollouts for $B_0 = [0.08,0]$.}
\label{tab:metrics_comparison_B0_08}
\resizebox{\textwidth}{!}{%
\begin{tabular}{lcccccccc}
\hline
Model & Quantities & MSE & RMSE & MAE & SSIM & PSNR (dB) & Epochs \\
\hline
KT-MHD & ($\omega, j$) & (0.095, 6.78) & (0.30, 2.60) & (0.11, 0.48) & (0.72, 0.58) & (32.27, 25.45) & 500 \\
ConvLSTM-UNet & ($\omega, j$) & (0.085, 6.44) & (0.29, 2.53) & (0.10, 0.47) & (0.74, 0.62) & (33.57, 25.84) & 500 \\
\hline
\end{tabular}}
\end{table*}

\begin{table}[h]
\centering
\caption{Global invariant conservation metrics for Koopman-Transformer and ConvLSTM-UNet models on the Kelvin-Helmholtz test set ($B_0=0.08$, horizon=2, rollout=6). All errors are averaged over all sequences.}
\label{tab:invariant_comparison_B0_0_08}
\begin{tabular}{lcc}
\hline\hline
\textbf{Metric} & \textbf{Transformer} & \textbf{LSTM-UNet} \\
\hline
Mean relative error $E_{\mathrm{tot}}$ (\%) & 9.86 & 6.31 \\

Mean energy drift (\%)                      & 13.28 & 1.98 \\
RMSE $H_c$ ($\times10^{-3}$)               & 2.04 & 5.39 \\
\hline\hline
\end{tabular}
\end{table}

%\nocite{*}
\bibliography{references}

@book{freidberg2014ideal,
  author = {J. Freidberg},
  title = {Ideal Magnetohydrodynamics},
  publisher = {Cambridge University Press},
  year = {2014},
  address = {Cambridge, UK}
}

@article{rosofsky2023magnetohydrodynamics,
  title={Magnetohydrodynamics with physics informed neural operators},
  author={Rosofsky, Shawn G and Huerta, Eliu A},
  journal={Machine Learning: Science and Technology},
  volume={4},
  number={3},
  pages={035002},
  year={2023},
  publisher={IOP Publishing}
}

@article{varennes2025robust,
  title={A robust data-driven approach for modeling turbulent transport},
  author={Varennes, R and Qu, ZS and Cho, YW and Wan, C and Li, K and Heinonen, RA and Grandgirard, V},
  journal={Nuclear Fusion},
  volume={65},
  number={7},
  pages={076027},
  year={2025},
  publisher={IOP Publishing}
}

@article{garrido2025ai,
  title={An {A}{I}-driven reduced order model for edge tokamak turbulence},
  author={Garrido Gonz{\'a}lez, David and Saura, Nathaniel and Beyer, Peter and Mouhali, Waleed and Benkadda, Sadruddin},
  journal={Physics of Plasmas},
  volume={32},
  number={9},
  year={2025},
  publisher={AIP Publishing}
}

@article{asaka2024machine,
  title={Machine learning-based vorticity evolution and super-resolution of homogeneous isotropic turbulence using wavelet projection},
  author={Asaka, Tomoki and Yoshimatsu, Katsunori and Schneider, Kai},
  journal={Physics of Fluids},
  volume={36},
  number={2},
  year={2024},
  publisher={AIP Publishing}
}

@article{hoshino2025estimation,
  title={Estimation of Electrostatic Potential Fluctuations in {K}elvin-{H}elmholtz Turbulence in Linear Plasmas using Multi-Scale CNN},
  author={Hoshino, Shuta and Sasaki, Makoto and Ishikawa, Ryohtaroh T and Nakata, Motoki},
  journal={Plasma and Fusion Research},
  volume={20},
  pages={1203035},
  year={2025},
  publisher={The Japan Society of Plasma Science and Nuclear Fusion Research}
}

@article{bormanis2024solving,
  title={Solving the {O}rszag--{T}ang vortex magnetohydrodynamics problem with physics-constrained convolutional neural networks},
  author={Bormanis, A and Leon, Christopher Anders and Scheinker, Alexander},
  journal={Physics of Plasmas},
  volume={31},
  number={1},
  year={2024},
  publisher={AIP Publishing}
}

@article{patil2023autoregressive,
  title={Autoregressive transformers for data-driven spatiotemporal learning of turbulent flows},
  author={Patil, Aakash and Viquerat, Jonathan and Hachem, Elie},
  journal={APL Machine Learning},
  volume={1},
  number={4},
  year={2023},
  publisher={AIP Publishing}
}

@article{wu2025physics,
  title={Physics-informed neural networks for {K}elvin--{H}elmholtz instability with spatiotemporal and magnitude multiscale},
  author={Wu, Jiahao and Wu, Yuxin and Li, Xin and Zhang, Guihua},
  journal={Physics of Fluids},
  volume={37},
  number={3},
  year={2025},
  publisher={AIP Publishing}
}

@book{biskamp2003magnetohydrodynamic,
  title={Magnetohydrodynamic turbulence},
  author={Biskamp, Dieter},
  year={2003},
  publisher={Cambridge University Press}
}

@article{Miura1984KHI,
  title={Anomalous transport by magnetohydrodynamic {K}elvin-{H}elmholtz instabilities in the solar wind},
  author={Miura, Akira},
  journal={Journal of Geophysical Research},
  volume={89},
  number={A2},
  pages={801--818},
  year={1984},
  doi={10.1029/JA089iA02p00801}
}

@article{Tirunagari2015KH,
  title={Exploratory data analysis of the {Ke}lvin-{H}elmholtz instability in jets},
  author={Tirunagari, Santosh},
  journal={arXiv preprint arXiv:1503.06331},
  year={2015}
}

@article{yin2024characteristic,
  title={A Characteristic Mapping Method with Source Terms: Applications to Ideal Magnetohydrodynamics},
  author={Yin, Xi-Yuan and Krah, Philipp and Nave, Jean-Christophe and Schneider, Kai},
  journal={Journal of Computational Physics},
  volume={555},
  pages={114766},
  year={2026}
}

@article{lusch2018deep,
  title={Deep learning for universal linear embeddings of nonlinear dynamics},
  author={Lusch, Bethany and Kutz, J Nathan and Brunton, Steven L},
  journal={Nature communications},
  volume={9},
  number={1},
  pages={4950},
  year={2018},
  publisher={Nature Publishing Group UK London}
}

@article{li2024deep,
  title={A deep {U}-{N}et-conv{LSTM} framework with hydrodynamic model for basin-scale hydrodynamic prediction},
  author={Li, Ao and Zhang, Wanshun and Zhang, Xiao and Chen, Gang and Liu, Xin and Jiang, Anna and Zhou, Feng and Peng, Hong},
  journal={Water},
  volume={16},
  number={5},
  pages={625},
  year={2024},
  publisher={Multidisciplinary Digital Publishing Institute}
}

@article{vaswani2017attention,
  title={Attention is all you need},
  author={Vaswani, Ashish and Shazeer, Noam and Parmar, Niki and Uszkoreit, Jakob and Jones, Llion and Gomez, Aidan N and Kaiser, {\L}ukasz and Polosukhin, Illia},
  journal={Advances in neural information processing systems},
  volume={30},
  year={2017}
}

@article{karniadakis2021physics,
  title={Physics-informed machine learning},
  author={Karniadakis, George Em and Kevrekidis, Ioannis G and Lu, Lu and Perdikaris, Paris and Wang, Sifan and Yang, Liu},
  journal={Nature Reviews Physics},
  volume={3},
  number={6},
  pages={422--440},
  year={2021},
  publisher={Nature Publishing Group UK London}
}

@inproceedings{akiba2019optuna,
  title={Optuna: A next-generation hyperparameter optimization framework},
  author={Akiba, Takuya and Sano, Shotaro and Yanase, Toshihiko and Ohta, Takeru and Koyama, Masanori},
  booktitle={Proceedings of the 25th ACM SIGKDD international conference on knowledge discovery \& data mining},
  pages={2623--2631},
  year={2019}
}

@inproceedings{ronneberger2015u,
  title={U-net: Convolutional networks for biomedical image segmentation},
  author={Ronneberger, Olaf and Fischer, Philipp and Brox, Thomas},
  booktitle={International Conference on Medical image computing and computer-assisted intervention},
  pages={234--241},
  year={2015},
  organization={Springer}
}

@article{vlachas2018data,
  title={Data-driven forecasting of high-dimensional chaotic systems with long short-term memory networks},
  author={Vlachas, Pantelis R and Byeon, Wonmin and Wan, Zhong Y and Sapsis, Themistoklis P and Koumoutsakos, Petros},
  journal={Proceedings of the Royal Society A: Mathematical, Physical and Engineering Sciences},
  volume={474},
  number={2213},
  year={2018},
  publisher={The Royal Society}
}

@article{brunton2019notes,
  title={Notes on {K}oopman operator theory},
  author={Brunton, Steven L},
  journal={Universit{\"a}t von Washington, Department of Mechanical Engineering, Zugriff},
  volume={30},
  year={2019}
}

@article{kraichnan1967inertial,
  title={Inertial ranges in two-dimensional turbulence},
  author={Kraichnan, Robert H.},
  journal={Physics of Fluids},
  volume={10},
  number={7},
  pages={1417--1423},
  year={1967},
  doi={10.1063/1.1762301}
}

@article{frisch1975possibility,
  title={Possibility of an inverse cascade of magnetic helicity in magnetohydrodynamic turbulence},
  author={Frisch, Uriel and Pouquet, A and L{\'e}orat, J and Mazure, A},
  journal={Journal of Fluid Mechanics},
  volume={68},
  number={4},
  pages={769--778},
  year={1975},
  publisher={Cambridge University Press}
}

@book{davidson2017introduction,
  title={Introduction to magnetohydrodynamics},
  author={Davidson, Peter Alan},
  year={2017},
  publisher={Cambridge university press}
}

@article{chen2026convolution,
  title={Convolution Operator Network for Forward and Inverse Problems ({FI}-Conv): Application to Plasma Turbulence Simulations},
  author={Chen, Xingzhuo and Poole, Anthony and Farcas, Ionut-Gabriel and Hatch, David R and Braga-Neto, Ulisses},
  journal={arXiv preprint arXiv:2602.04287},
  year={2026}
}

@article{constante2024data,
  title={Data-driven {K}oopman operator predictions of turbulent dynamics in models of shear flows},
  author={Constante-Amores, C Ricardo and Fox, Andrew J and De Jes{\'u}s, Carlos E P{\'e}rez and Graham, Michael D},
  journal={arXiv preprint arXiv:2407.16542},
  year={2024}
}

@book{brunton2022data,
  title={Data-Driven Science and Engineering: Machine Learning, Dynamical Systems, and Control},
  author={Brunton, Steven L. and Kutz, J. Nathan},
  year={2022},
  edition={2},
  publisher={Cambridge University Press},
  doi={10.1017/9781009089517}
}

@book{courant2024methods,
  title={Methods of mathematical physics, volume 2},
  author={Courant, Richard and Hilbert, David},
  year={2024},
  publisher={John Wiley \& Sons}
}

@article{faganello2012magnetic,
  title={Magnetic reconnection and {K}elvin--{H}elmholtz instabilities at the Earth's magnetopause},
  author={Faganello, Matteo and Califano, Francesco and Pegoraro, Francesco and Andreussi, T and Benkadda, S},
  journal={Plasma Physics and Controlled Fusion},
  volume={54},
  number={12},
  pages={124037},
  year={2012},
  publisher={IOP Publishing}
}

@article{long2025stft,
  title={St{FT}: Spatio-temporal {F}ourier Transformer for Long-term Dynamics Prediction},
  author={Long, Da and Zhe, Shandian and Williams, Samuel and Oliker, Leonid and Bai, Zhe},
  journal={arXiv preprint arXiv:2503.11899},
  year={2025}
}

@article{brunton2016koopman,
  title={Koopman invariant subspaces and finite linear representations of nonlinear dynamical systems for control},
  author={Brunton, Steven L and Brunton, Bingni W and Proctor, Joshua L and Kutz, J Nathan},
  journal={PloS one},
  volume={11},
  number={2},
  pages={e0150171},
  year={2016},
  publisher={Public Library of Science San Francisco, CA USA}
}

@article{lin2024synthesizing,
  title={Synthesizing impurity clustering in the edge plasma of tokamaks using neural networks},
  author={Lin, Zetao and Maurel-Oujia, Thibault and Kadoch, Benjamin and Krah, Philipp and Saura, Nathaniel and Benkadda, Sadruddin and Schneider, Kai},
  journal={Physics of Plasmas},
  volume={31},
  number={3},
  year={2024},
  publisher={AIP Publishing}
}

@inproceedings{desai2022next,
  title={Next frame prediction using Conv{LSTM}},
  author={Desai, Padmashree and Sujatha, C and Chakraborty, Saumyajit and Ansuman, Saurav and Bhandari, Sanika and Kardiguddi, Sharan},
  booktitle={Journal of Physics: Conference Series},
  volume={2161},
  number={1},
  pages={012024},
  year={2022},
  organization={IOP Publishing}
}

@article{wang2004image,
  title={Image quality assessment: from error visibility to structural similarity},
  author={Wang, Zhou and Bovik, Alan C and Sheikh, Hamid R and Simoncelli, Eero P},
  journal={IEEE transactions on image processing},
  volume={13},
  number={4},
  pages={600--612},
  year={2004},
  publisher={IEEE}
}

@book{chandrasekhar2013hydrodynamic,
title={Hydrodynamic and hydromagnetic stability},
author={Chandrasekhar, Subrahmanyan},
year={2013},
publisher={Courier Corporation}
}

@article{courant1928partiellen,
  title={{\"U}ber die partiellen {D}ifferenzengleichungen der mathematischen {P}hysik},
  author={Courant, Richard and Friedrichs, Kurt and Lewy, Hans},
  journal={Mathematische Annalen},
  volume={100},
  number={1},
  pages={32--74},
  year={1928},
  publisher={Springer}
}

\end{document}